\documentclass[%
aip,
jcp,%
preprint,%
groupedaddress
]{revtex4-1}

\usepackage{graphicx}
\usepackage{dcolumn}
\usepackage{bm}
\usepackage{amsmath}
\usepackage{setspace}
\usepackage{color}

\usepackage{float}

\usepackage{array}
\usepackage{multirow}
\usepackage{tabulary}
\usepackage{subfigure}

\begin{document}


\title[]{Heterogeneous dynamics and its length scale in simple ionic liquid models: A computational study}

\author{Soree Kim}
\affiliation{Department of Chemistry, Seoul National University, Seoul 08826, Korea}

\author{Sang-Won Park}%
\affiliation{Department of Chemistry, Seoul National University, Seoul 08826, Korea}

\author{YounJoon Jung}
\email{yjjung@snu.ac.kr}
\affiliation{Department of Chemistry, Seoul National University, Seoul 08826, Korea}

\date{\today}

\begin{abstract}
We numerically investigate the dynamic heterogeneity and its length scale found in the coarse-grained ionic liquid model systems.
In our ionic liquid model systems, cations are modeled as dimers with positive charge, while anions are modeled as monomers with negative charge, respectively.
To study the effect of the charge distributions on the cations,
two ionic liquid models with different charge distributions are used and the model with neutral charge is also considered as a counterpart.
To reveal the heterogeneous dynamics in the model systems,
we examine spatial distributions of displacement and time distributions of exchange and persistence times.
All the models show significant increase of the dynamic heterogeneity as the temperature is lowered.
The dynamic heterogeneity is quantified via the well-known four-point susceptibility, $\chi_4(t)$,
which measures the fluctuation of a time correlation function.
The dynamic correlation length is calculated by fitting the dynamic structure factor, $S_4(k,t)$,
with Ornstein-Zernike form at the time scale at which the dynamic heterogeneity reaches the maximum value.
Obtained time and length scales exhibit a power law relation at the low temperatures, similar to various supercooled liquid models.
Especially, the charged model systems show unusual crossover behaviors which are not observed in the uncharged model system.
We ascribe the crossover behavior to the enhanced cage effect caused by charges on the particles.
\end{abstract}

\maketitle


\section{Introduction}
Room-temperature ionic liquids (RTILs) have attracted great attention because of their uncommon physical properties and various applications including non-toxic solvents, electrolytes, and supercapacitors.\cite{holbrey1999ionic,wasserscheid2000ionic,weingartner2008understanding,deyoung2014graphene,shim2012graphene}
Widely known features of RTILs are their thermal stability, high polarity, high viscosity, very low vapor pressure, and low combustibility.
\cite{weingartner2008understanding}
Usually, RTILs are composed of bulky, asymmetric cations and small, symmetric anions.
Due to their considerable size difference, RTILs exist in a liquid phase near the room temperature in spite of the presence of strong Coulomb interaction.
One of the intriguing features of RTILs is their heterogeneous dynamics.
As reported by theoretical\cite{park2015lifetime,jeong2010fragility,del2004structure,wang2007understanding,kim2014dynamic}
 and experimental studies,\cite{russina2012mesoscopic,guo2011fluorescence,hideaki2005physical}
 the evidences of the heterogeneous dynamics such as a non-exponential decay of correlation functions have been found.
Computer simulation studies also have found the glassy dynamics of RTILs which is characterized by the breakdown of the Stokes-Einstein relation and decoupling of exchange and persistence events of defined excitations.\cite{park2015lifetime,jeong2010fragility}

There have been previous computational studies on the description of the heterogeneous dynamics in the ionic liquid systems.
\cite{urahata2004structure,wang2005unique,habasaki2008heterogeneous,hu2006heterogeneity,bhargava2007nanoscale}
However, the length scale of the dynamic heterogeneity has not been investigated thoroughly\cite{pal2014slow}
because of the difficulties on performing massive simulation with complex structures and long range interactions.
To overcome this difficulty and to enhance computational efficiency,
various levels of the coarse-grained models have been proposed.\cite{malvaldi2008molten,roy2010dynamics,del2004structure,wang2007understanding,jeong2010fragility}
Among these models, we use simple models of RTILs which are introduced by a previous study\cite{malvaldi2008molten}
and investigated intensively by Park et al..\cite{park2015lifetime} 

In the previous study, Ref \citenum{park2015lifetime}, 
we examined the structural and dynamic properties of RTILs thoroughly.
Especially, the relation between the time scale of the heterogeneous dynamics and the length scale of the structural relaxation were studied.
According to the simulation results, calculated lifetime of the heterogeneous dynamics, ${\tau}_{\text{dh}}$, could be regarded as a distinctive time scale from the relaxation time. 
While the lifetime of the heterogeneous dynamics is calculated using the three-time correlation functions,
the length scale was not examined.
In the present study, we further investigate the heterogeneous dynamics of RTILs initiated in Ref \citenum{park2015lifetime}.
We show the simulation results that support the existence of the heterogeneous dynamics in our model systems.
Furthermore, we present the simulation results on the length scale of the heterogeneous dynamics and the scaling relation between the relaxation time to show the distinctive nature of RTILs against the model without Coulomb interaction.

The heterogeneous dynamics found in RTILs have similar aspects found in the supercooled liquids. 
When the liquids are cooled down rapidly, they exist in supercooled liquids rather than forming a crystal structure. 
The viscosity and the structural relaxation time grow dramatically as the temperature of the system is lowered. 
The complete understanding of these physical phenomena and their theoretical explanations are still lacking. 
Among the distinctive behaviors of the supercooled liquids, the correlations between the time-dependent local density fluctuations are found to play an important role in the slowing down of system. 
This phenomenon, typically called dynamic heterogeneity, has been investigated through diverse theoretical
\cite{park2015lifetime,choi2015dynamic,kim2014dynamic,jung2005dynamical,jung2004excitation,pan2005heterogeneity,charbonneau2013decorrelation,karmakar2014growing,berthier2011theoretical,berthier2005length,berthier2004time,kim2013dynamic,lacevic2003spatially,lavcevic2002growing,glotzer2000time,toninelli2005dynamical,stein2008scaling,hedges2007decoupling}
 and experimental studies.\cite{berthier2005direct,wang1999long,cicerone1995relaxation,schmidt1991nature,heuer1995rate,ediger2000spatially}

In previous theoretical and computational studies, the time and length scales of the dynamic heterogeneity have been obtained using the four-point density correlation functions, \cite{dasgupta1991there,chandler2006lengthscale,flenner2014universal,flenner2013dynamic,flenner2011analysis,flenner2010dynamic,flenner2009anisotropic,kim2013dynamic,lacevic2003spatially,lavcevic2002growing,glotzer2000time,toninelli2005dynamical}
which has its origin in the study of spin glasses.\cite{young1997spin}
A four-point correlation function is defined as
\begin{equation}
\label{four-point}
\begin{split}
g_4(\mathbf{r},t)=\langle \delta\rho(0,0)\delta\rho(0,t)\delta\rho(\mathbf{r},0)\delta\rho(\mathbf{r},t) \rangle \\
-\langle \delta\rho(0,0)\delta\rho(0,t)\rangle\langle\delta\rho(\mathbf{r},0)\delta\rho(\mathbf{r},t) \rangle,
\end{split}
\end{equation}
where $\delta\rho(\mathbf{r},t)$ is the deviation of the local density at the position $\mathbf{r}$ and at time $t$.
$g_4(\mathbf{r},t)$ measures the correlation of relaxation of the density fluctuation between the two points separated by $\mathbf{r}$.
The dynamic susceptibility and the dynamic structure factor can be derived from this function,
by integrating and by performing Fourier transform, respectively.
We use the dynamic susceptibility as an index of the quantification of the dynamic heterogeneity and
define the time value that makes the dynamic susceptibility maximum as a characteristic time scale of the dynamic heterogeneity.
The dynamic structure factor is also calculated to extract the dynamic length scale, $\xi_4(t)$,
which could be interpreted as a length scale of dynamically correlated regions.
With these schemes, we find the characteristic time scale and the length scale of the dynamic heterogeneity in the ionic liquid model systems.

The contents of this article are organized as follows:
In Section 2, we introduce ionic liquid models and describe shortly the simulation methods.
In Section 3, the evidences of heterogeneous dynamics in the ionic liquid models are shown using displacement distributions and the decoupling of the mean exchange time and the mean persistence time.
Furthermore, the dynamic length scale obtained by calculating the four-point correlation functions and related scaling behavior will be illustrated.
Finally, the conclusions on this work are shown in the Section 4.

\section{Models}

\begin{figure}[h]
	\begin{center}
	\mbox{
		\subfigure{
			\includegraphics[width=0.4\textwidth]{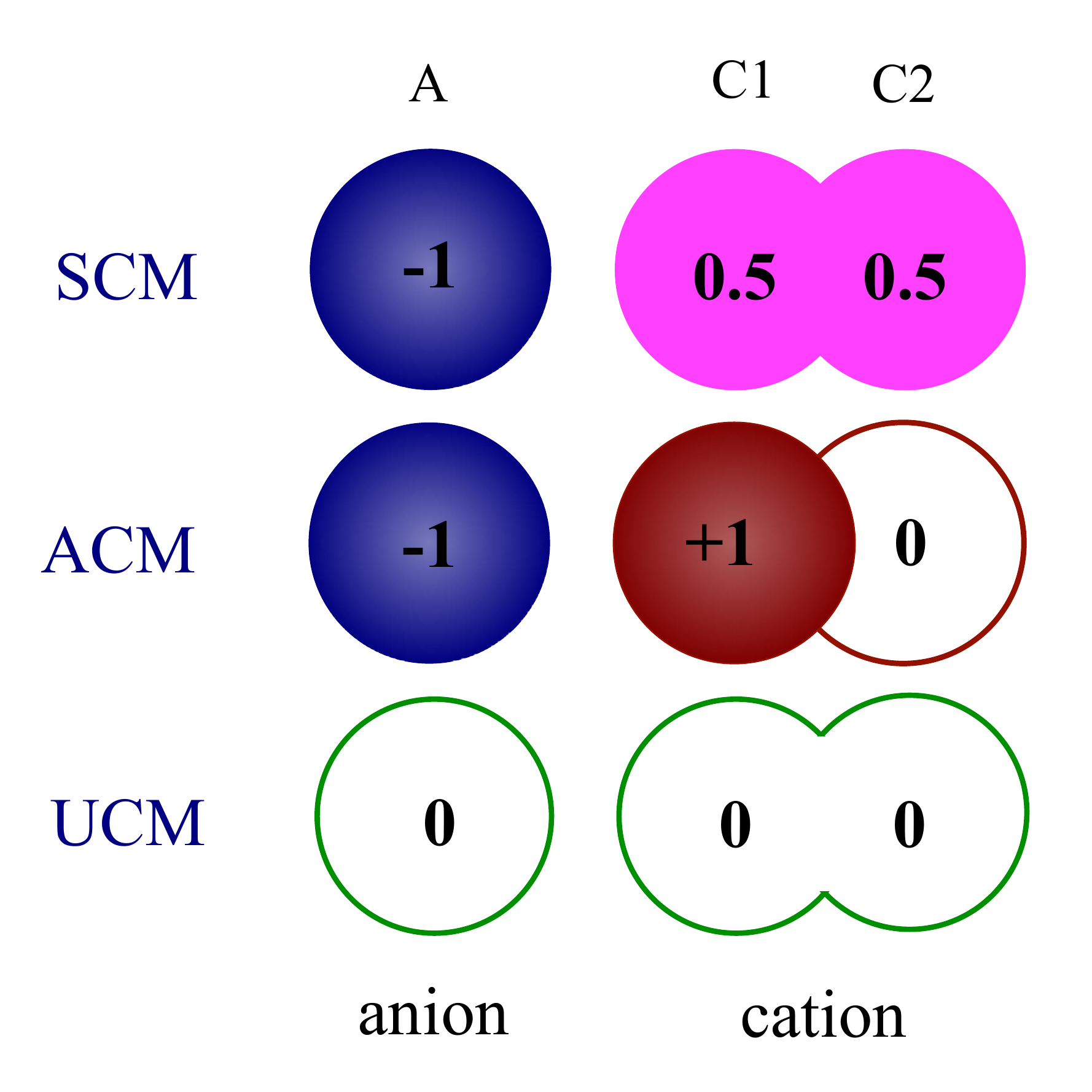}
			}	
		}
	\caption{
Schematic representation of the three models. 
SCM (top) and ACM (middle) represent symmetrically charged and asymmetrically charged model of ionic liquid, respectively.
While the anions of both model have the charge of $-1.0e$,
the cations have different charge distribution: 
the positive charge is equally distributed for the SCM cation particles ($+0.5e$ for C1 and C2),
the charge is separated for the ACM cation particles ($+1.0e$ for C1 and zero for C2).
UCM (bottom) denotes the uncharged model as a comparison group without charge on every particles.
Figure adapted from Ref.\citenum{park2015lifetime} with permission.
	}
	\label{model_new}	
	\end{center}
\end{figure}

We use simple coarse-grained models to investigate the heterogeneous dynamics of the room-temperature ionic liquid systems.
In order to study the effect of charge distribution on the cation, 
the symmetrically charged model (SCM) and the asymmetrically charged model (ACM) are used.
In addition, the uncharged model (UCM) is also used as a comparison group.
Three model systems have the cation composed of two particles and the anion of single particle (Fig.\ref{model_new}).
All the physical parameters are the same for those models except the charge distribution.
SCM has $+0.5 e$ (where $e$ is the elementary charge) on each particle in the cation and $-1.0 e$ on the anion,
while ACM has $+1.0 e$ on C1 particle, zero charge on C2 particle, and also $-1.0 e$ for the anion.
UCM has zero charge for all particles.
We also use the term ``cation'' and ``anion'' for the UCM, for convenience, even if the model does not have charges on the particles.  

\begin{figure*}[ht]
	\begin{center}
	\mbox{
	\hspace{-40pt}
		\subfigure[]{
			\includegraphics[width=0.4\textwidth]{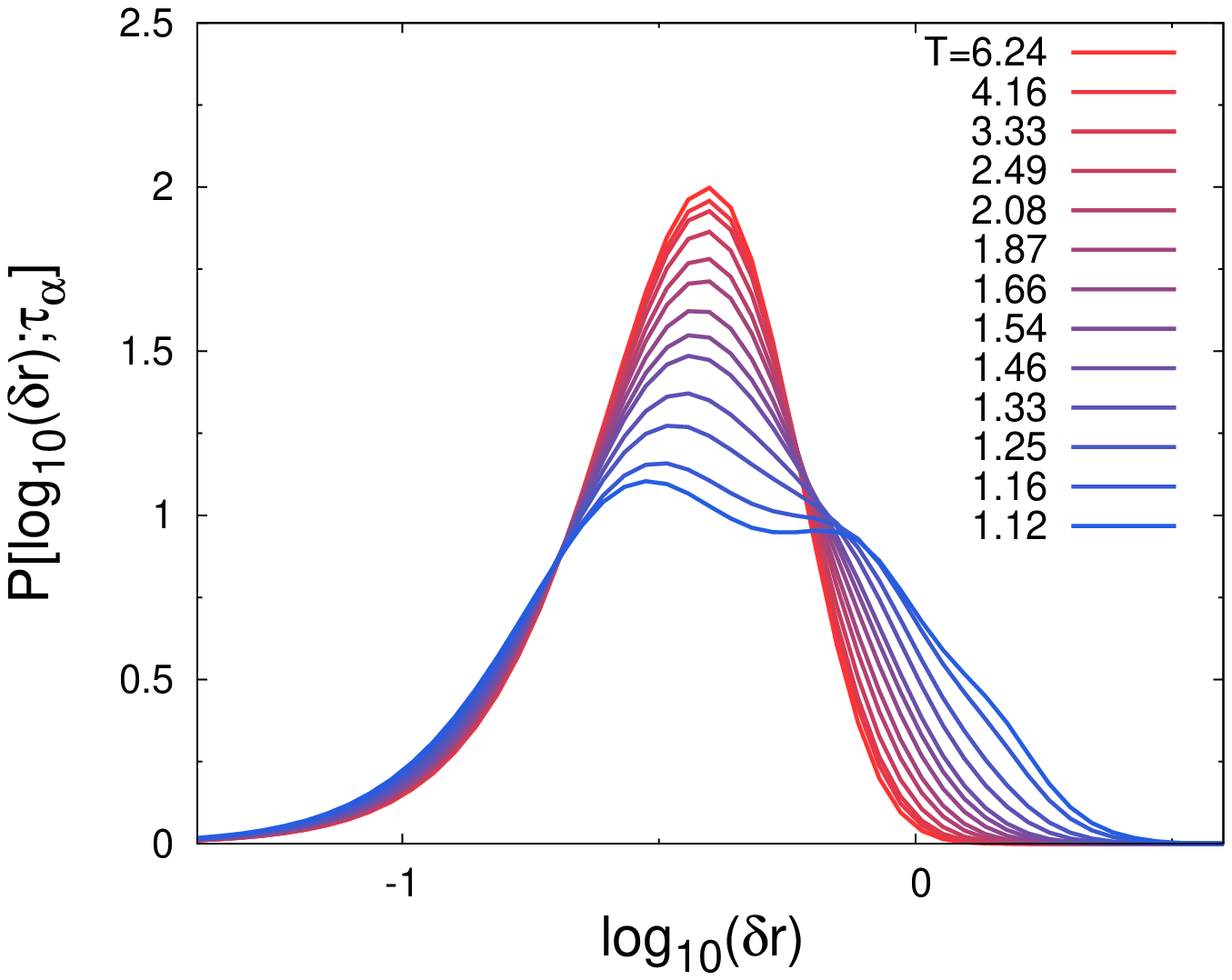}
		\label{dis_tscan_scm}			
			}
	\hspace{-36pt}
		\subfigure[]{
			\includegraphics[width=0.4\textwidth]{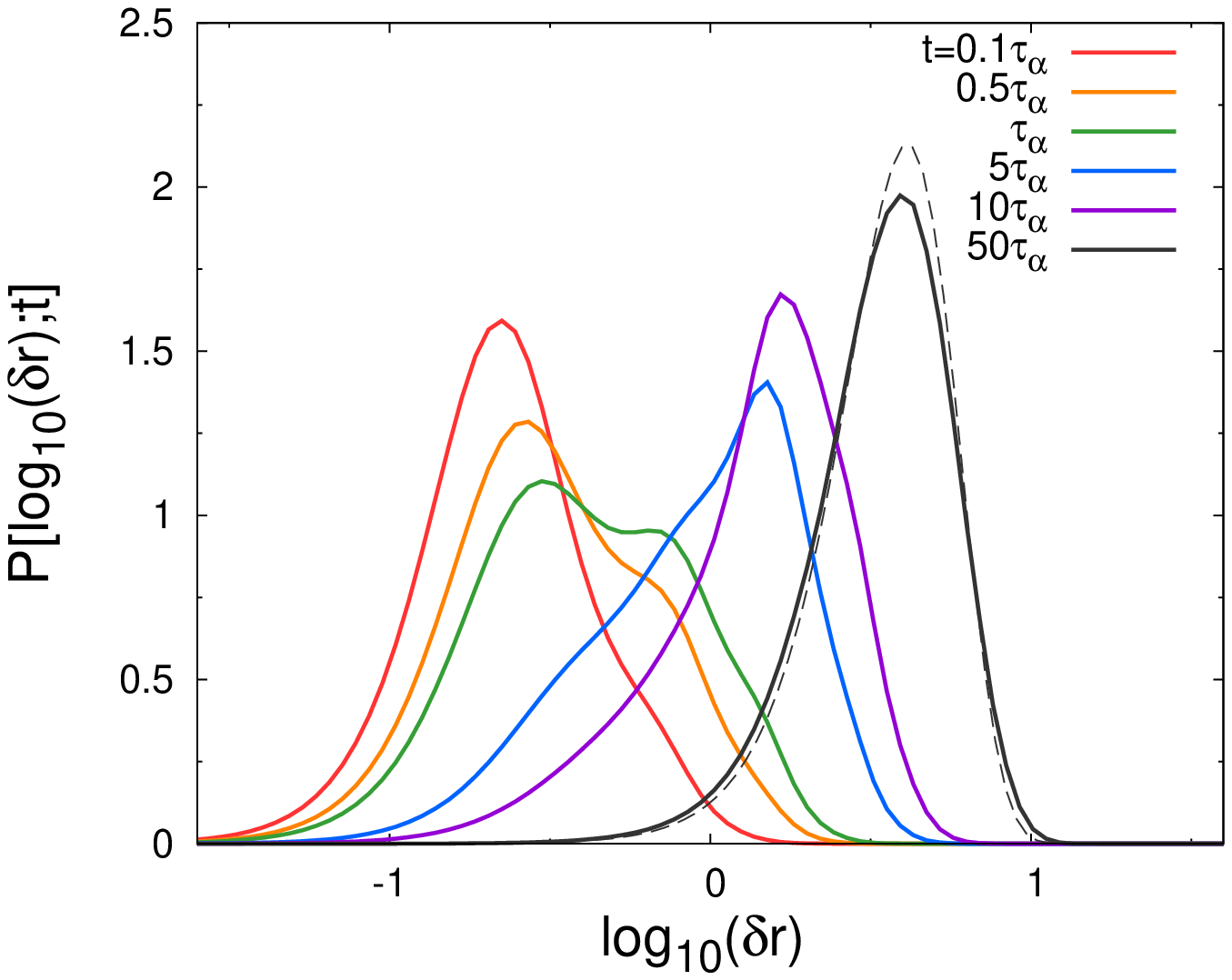}
			\label{dis_scm}
			}
	\hspace{-32pt}
		\subfigure[]{
			\includegraphics[width=0.4\textwidth]{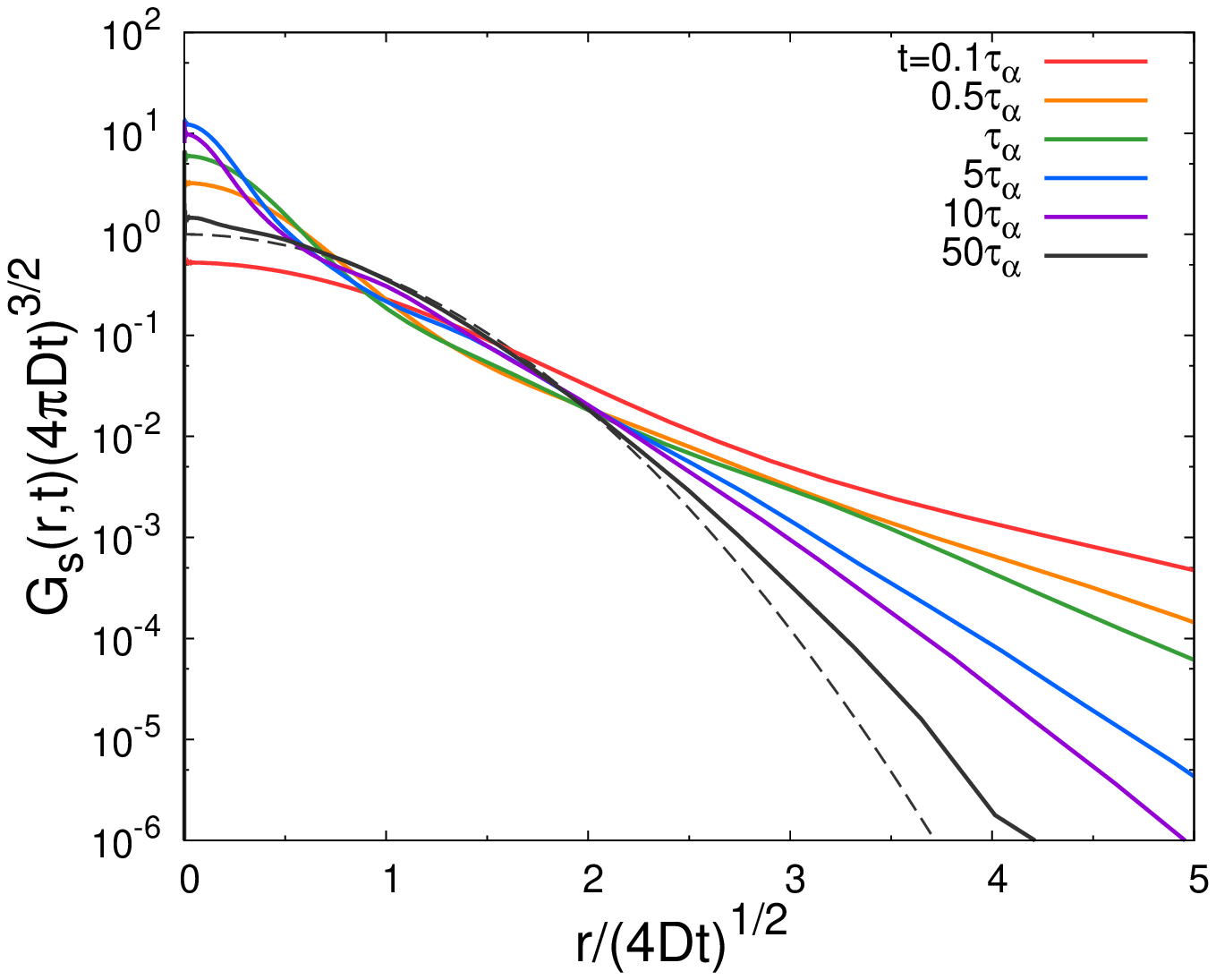}
			\label{g_s_cat}
			}
			}
	\mbox{
	\hspace{-40pt}
		\subfigure[]{
			\includegraphics[width=0.4\textwidth]{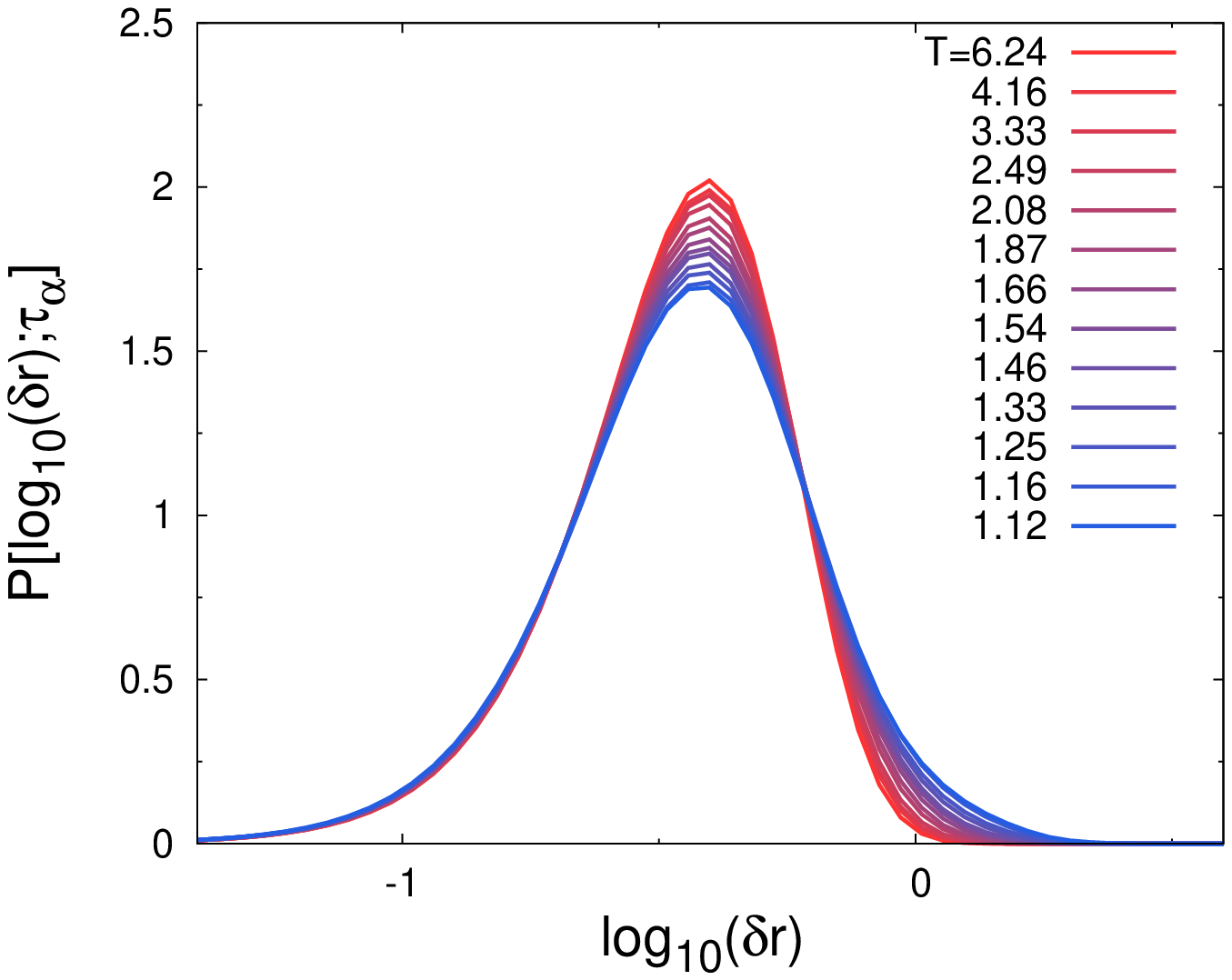}
		\label{dis_tscan_scm_ani}			
			}
	\hspace{-36pt}
		\subfigure[]{
			\includegraphics[width=0.4\textwidth]{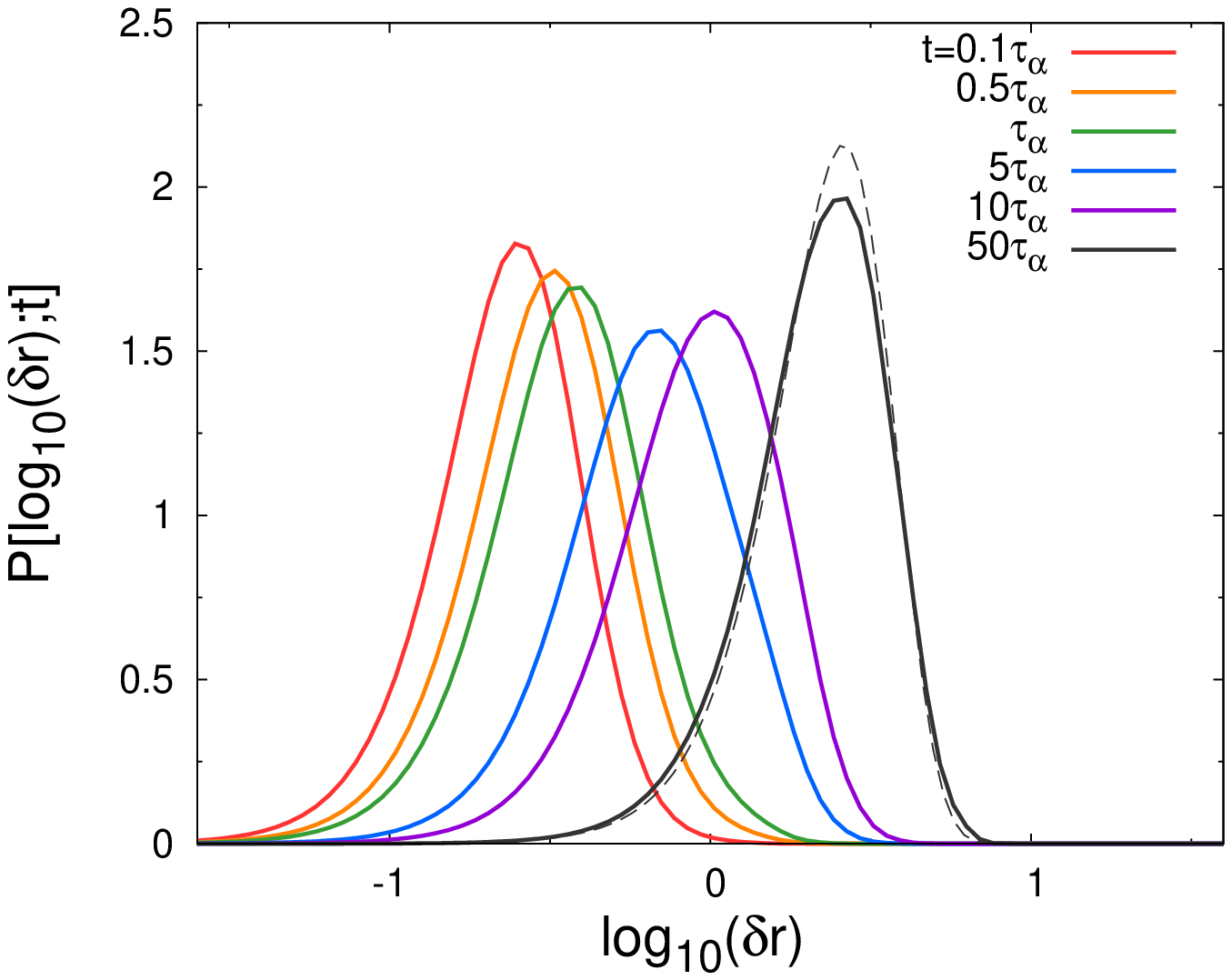}
			\label{dis_scm_ani}
			}
	\hspace{-32pt}
		\subfigure[]{
			\includegraphics[width=0.4\textwidth]{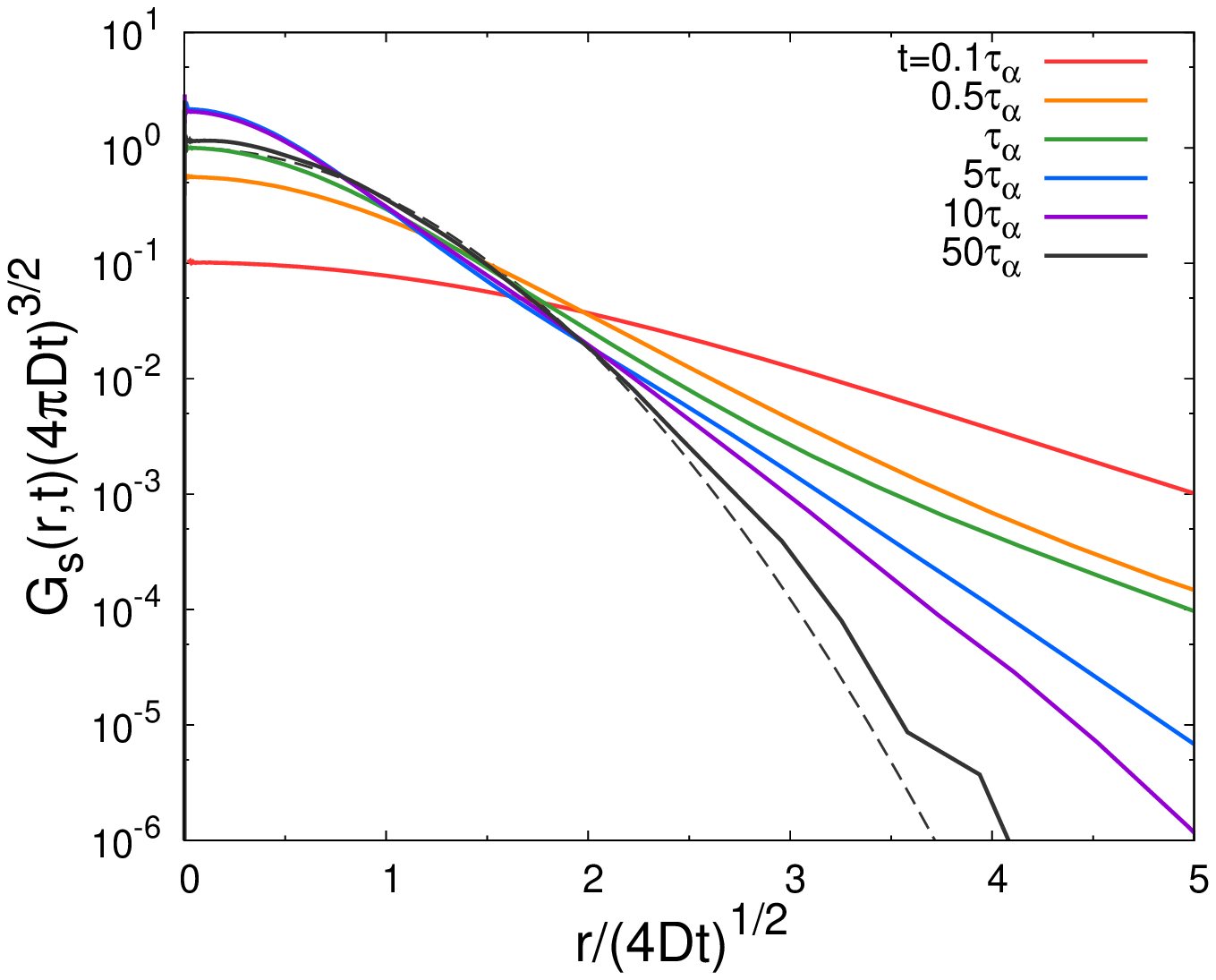}
			\label{g_s_ani}
			}
			}	
	\caption{
	Displacement distributions of (a) the cation and (d) the anion in SCM.
	As the temperature decreases, the displacement becomes heterogeneous.
	Time dependence of the displacement distributions of (b) the cation and (e) the anion in SCM at $T=1.12$.
	Both at short and long time cases, the distribution shows a single peak (dashed line shows Gaussian distribution).
	The distribution is heterogeneous at the time near the relaxation time $\tau_{\alpha}$.
	The self-van Hove functions of (c) the cation and (f) the anion are getting close to the Gaussian distribution (dashed line).
	However, there are still mismatches for the fast particles even in the long time limit.
	}
	\end{center}
	\label{displacement_dist}	
\end{figure*}

The total potential energy is given by the sum of the pairwise interactions,
\begin{equation}
U_{\text{total}} =\sum_{i,j}\{ U_{\text{LJ}}(r_{ij})+U_{\text{Coulomb}}(r_{ij}) \}
\end{equation}
where,
\begin{equation}
U_{\text{LJ}}(r_{ij}) =4\epsilon_{ij}\left\lbrack \left(\frac{\sigma_{ij}}{r_{ij}}\right)^{12}-\left(\frac{\sigma_{ij}}{r_{ij}}\right)^{6}+\frac{1}{4}\right\rbrack H(r_{\text{cut}}-r_{ij}),
\end{equation}
and
\begin{equation}
U_{\text{Coulomb}}(r_{ij}) = \frac{1}{4\pi\epsilon_0}\frac{q_i q_j e^2}{r_{ij}}.
\end{equation}
$H(r_{\text{cut}}-r_{ij})$ is the Heaviside step function, where the cutoff distance is set to be $r_{\text{cut}}=2^{1/6}\sigma_{ij}$. 
Note that $U_{\text{LJ}}(r_{ij})$ is purely repulsive and it is called the Weeks-Chandler-Andersen (WCA) potential.\cite{weeks1971role}
In all of the three models, $\epsilon_{ij}=\epsilon=2$ kJ/mol and $\sigma_{ij}= \sigma=0.5$ nm for all $i,j$ pairs.
The length of rigid bond between two particles (C1 and C2) in the cation is set to be $0.8\sigma$.
The mass of the particles in the cation is $m=100$ amu and the mass of the anion is 200 amu, so the total mass of the cation and that of the anion are the same.
We use the length, energy, and mass scaled by the units of $\sigma$, $\epsilon$, and $m$.
The other units are converted by the following relations: unit time, $t_0=(m\sigma^2/\epsilon)^{1/2}=5$ ps,
unit temperature, $T_0=\epsilon/k_{\text{B}}=240.5$ K, unit charge, $q_0=(4\pi \epsilon_0 \sigma \epsilon)^{1/2}=0.08484$ $e$, and unit pressure, $P_0 =262.2$ atm.
We use 2048 pairs of RTIL molecules in a cubic simulation box of $L=17.88$, where $L$ is the length of each side.
All the system have the reduced number density $\rho^{*}=\rho \sigma^3 =0.716$.

We perform molecular dynamics (MD) simulation using GROMACS 4.5 MD package program\cite{pronk2013gromacs}
 under $NVT$ ensemble condition with Nos$\acute{\text{e}}$-Hoover thermostat.
Periodic boundary condition is applied to each direction and the
finite size effect is carefully checked by comparing physical quantities calculated from systems with different size of 512, 1024, 2048 and 4096 RTIL pairs.
For all the systems, ten independent trajectories are used and the length of production run is about 40 times of
the $\alpha$-relaxation time of each system.
The details of the molecular dynamics simulation conditions are given in Ref.\citenum{park2015lifetime}.


\section{Results and discussion}

\subsection{Heterogeneous dynamics}
\begin{figure*}[!htb]
	\begin{center}
	\mbox{
	\hspace{-40pt}
		\subfigure[]{
			\includegraphics[width=0.4\textwidth]{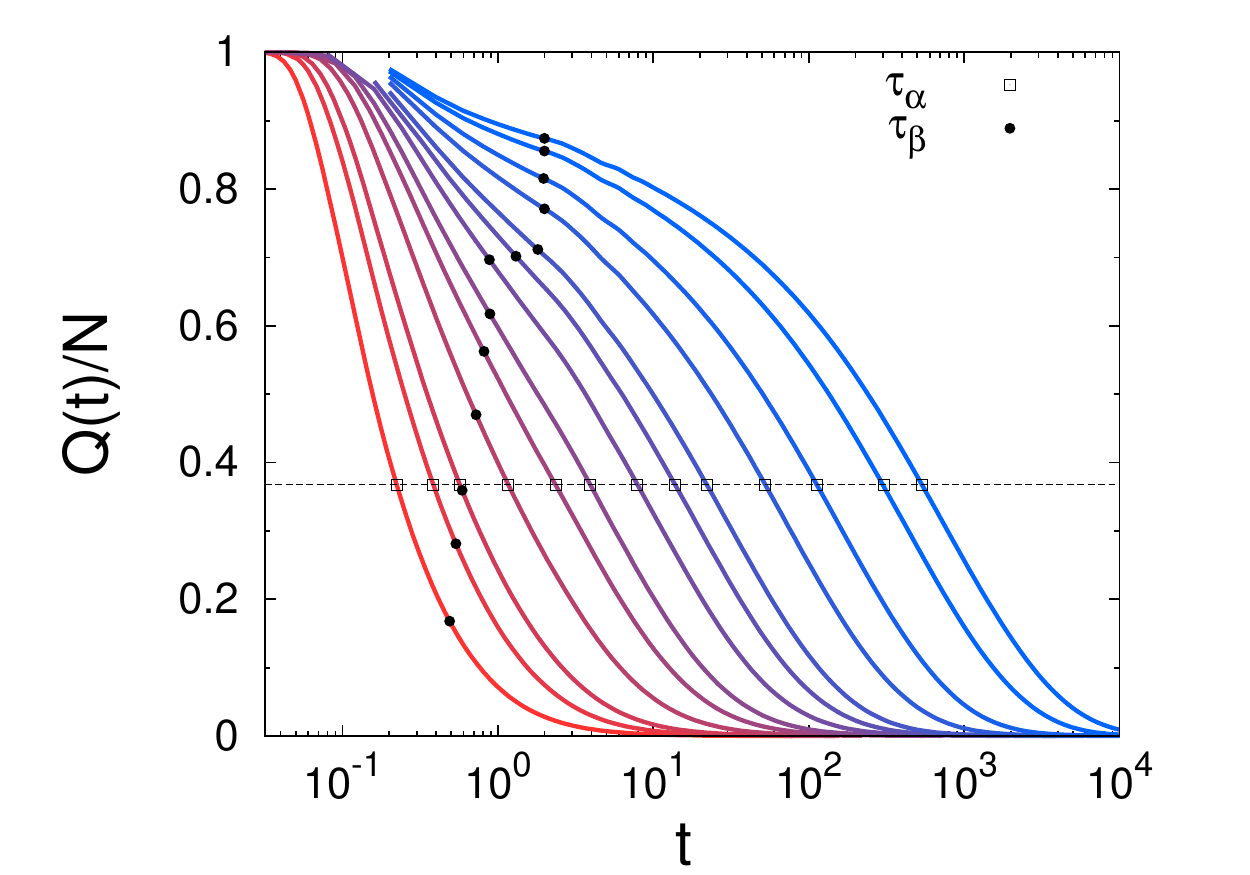}
			\label{q_t_scm}
			}
	\hspace{-36pt}
		\subfigure[]{
			\includegraphics[width=0.4\textwidth]{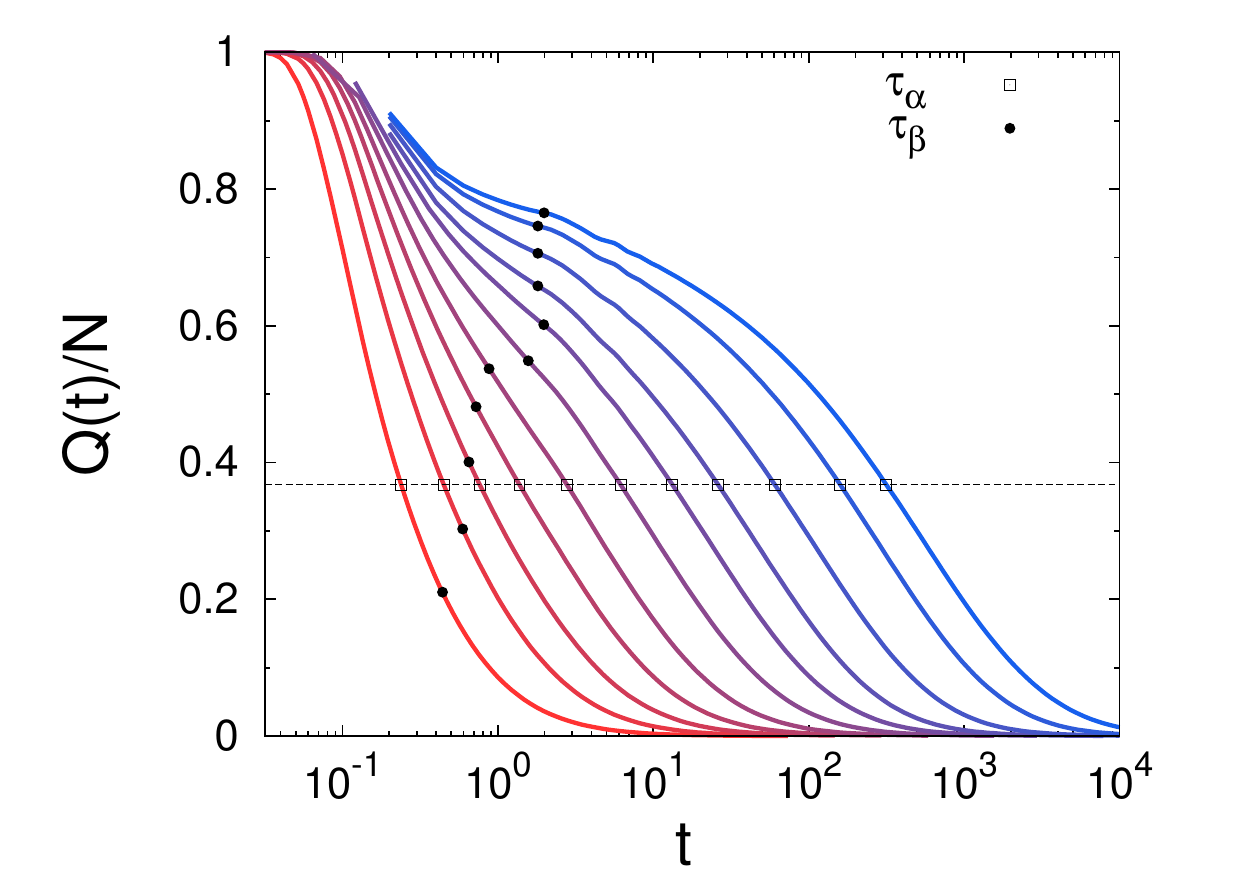}
			\label{q_t_acm}
		}
	\hspace{-32pt}
		\subfigure[]{
			\includegraphics[width=0.4\textwidth]{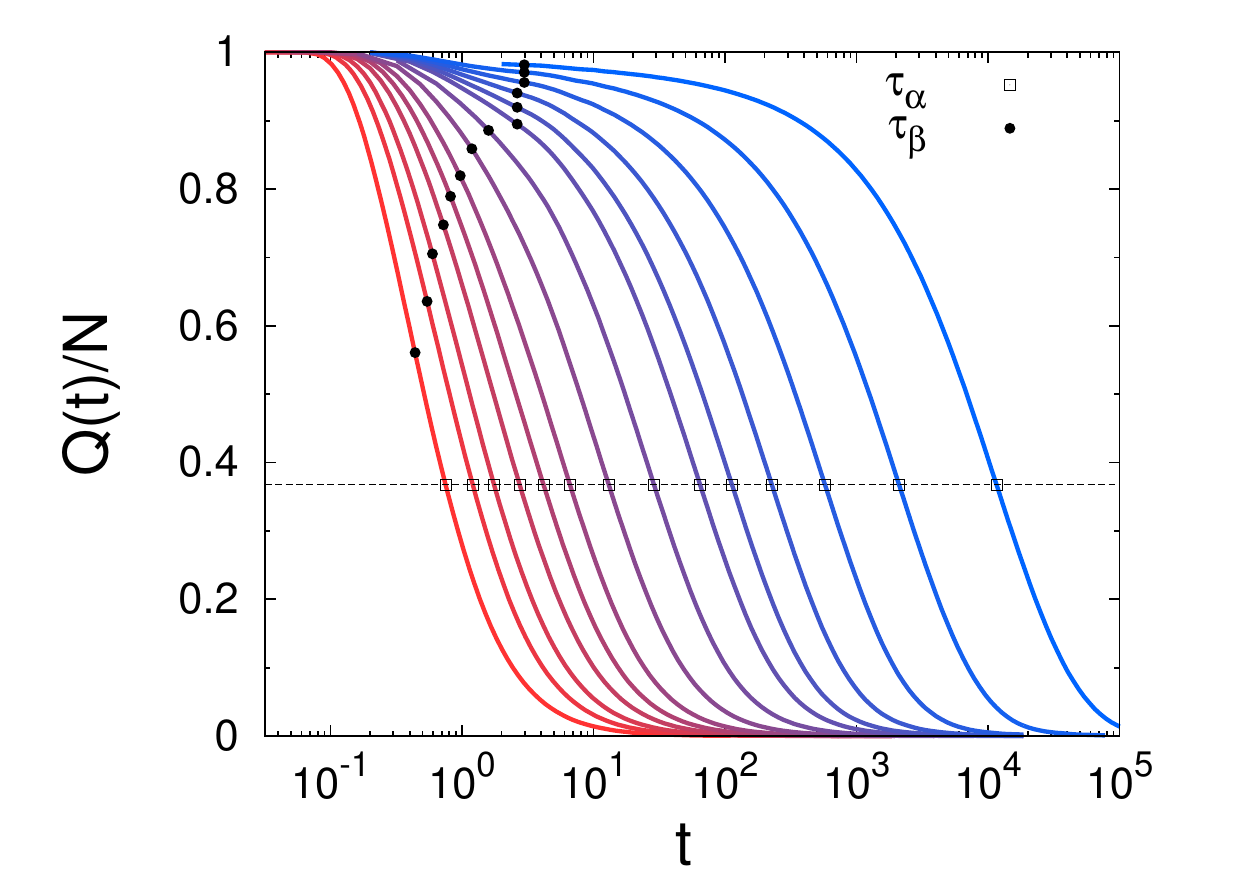}
			\label{q_t_ucm}
		}
		}
	\caption{
Normalized overlap function, $Q(t)/N$, of the cations in each system calculated using Eq.\ref{eq-q_t} at various temperatures (a)SCM, (b)ACM and (c)UCM:
from left to right,
$T$= 6.24, 4.16, 3.33, 2.49, 2.08, 1.87, 1.66, 1.54, 1.46, 1.33, 1.25, 1.16 and 1.12 (SCM),
$T$= 6.24, 4.16, 3.33, 2.83, 2.49, 2.25, 2.08, 1.98, 1.87, 1.79 and 1.75 (ACM),
$T$= 1.56, 1.14, 0.94, 0.77, 0.67, 0.58, 0.50, 0.44, 0.40, 0.37, 0.35, 0.33 0.31 and 0.29 (UCM).
The $\alpha$-relaxation time, $\tau_{\alpha}$, is defined at which $Q(\tau_{\alpha})/N=1/e$.
The $\beta$-relaxation time, $\tau_{\beta}$, is the characteristic time scale for the cage effect (See Fig.\ref{cal_tau_beta}).
	}
	\label{q_t}
	\end{center}
\end{figure*}

Our ionic liquid model systems are expected to have the heterogeneous dynamics because of the size difference between the cation and the anion.
To investigate the heterogeneous dynamics in detail,
we first calculate the displacement of each particle. 
Fig.\ref{dis_tscan_scm} shows the probability of the logarithm of displacements, $P[\text{log}_{10}(\delta r);t]$, of the cation in SCM.
The time $t$ is set to be the $\alpha$-relaxation time, $t=\tau_{\alpha}$, at each temperature,
where the $\alpha$-relaxation time is defined by the time at which the normalized overlap function, $Q(t)/N$, falls into $1/e$ (See Fig.\ref{q_t}).
The definition of the overlap function will be introduced in Section 3.2.
$P[\text{log}_{10}(\delta r);t]$ is related to the self-van Hove function, $G_s(\delta r;t)$,
through the equation, $P[\text{log}_{10}(\delta r);t]=4\pi \text{log}(10){\delta r}^3 G_s(\delta r;t)$.
Since $G_s(\delta r;t)$ follows a Gaussian function when the particle experienced the Fickian diffusion,
$P[\text{log}_{10}(\delta r);t]$ would show a single peak.
Therefore, the broadening or split of the distribution is a clear evidence for non-Fickian motion and heterogeneous dynamics.

\begin{figure*}[ht]
	\begin{center}
	\mbox{
		\subfigure[]{
			\includegraphics[width=0.45\textwidth]{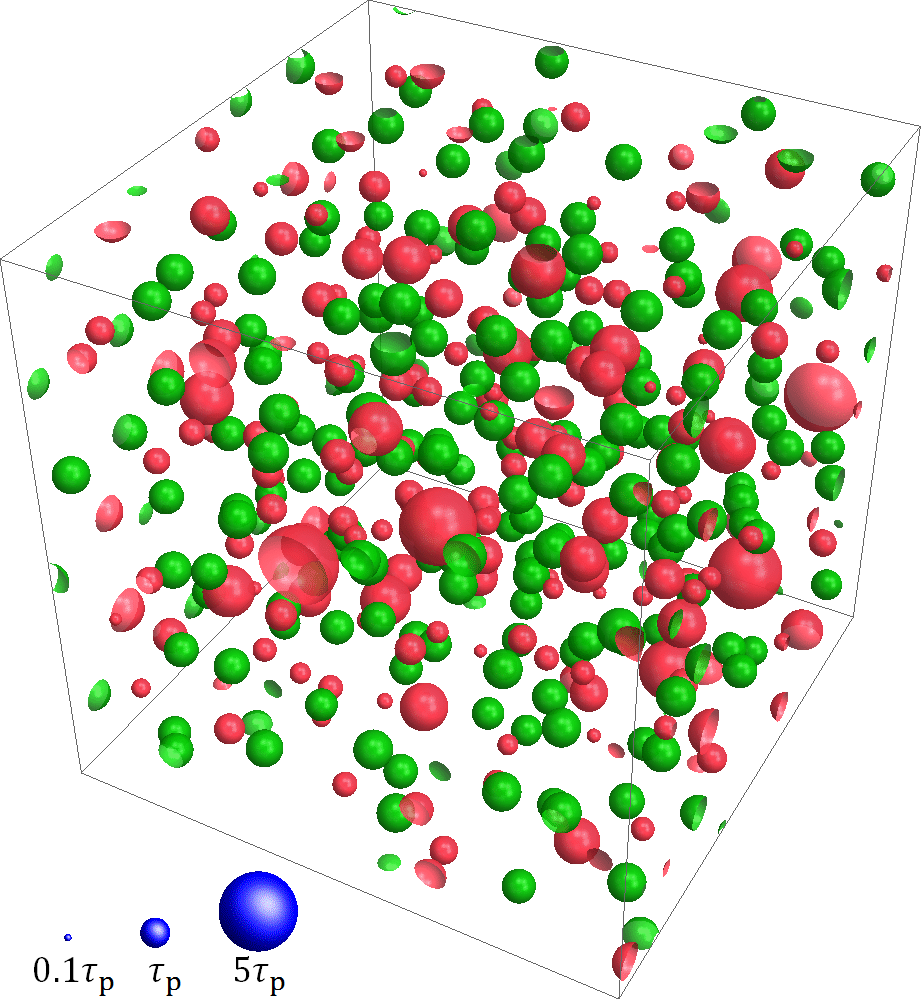}
			\label{het_ca_scm_high}
			}
			\hspace{20pt}
		\subfigure[]{
			\includegraphics[width=0.45\textwidth]{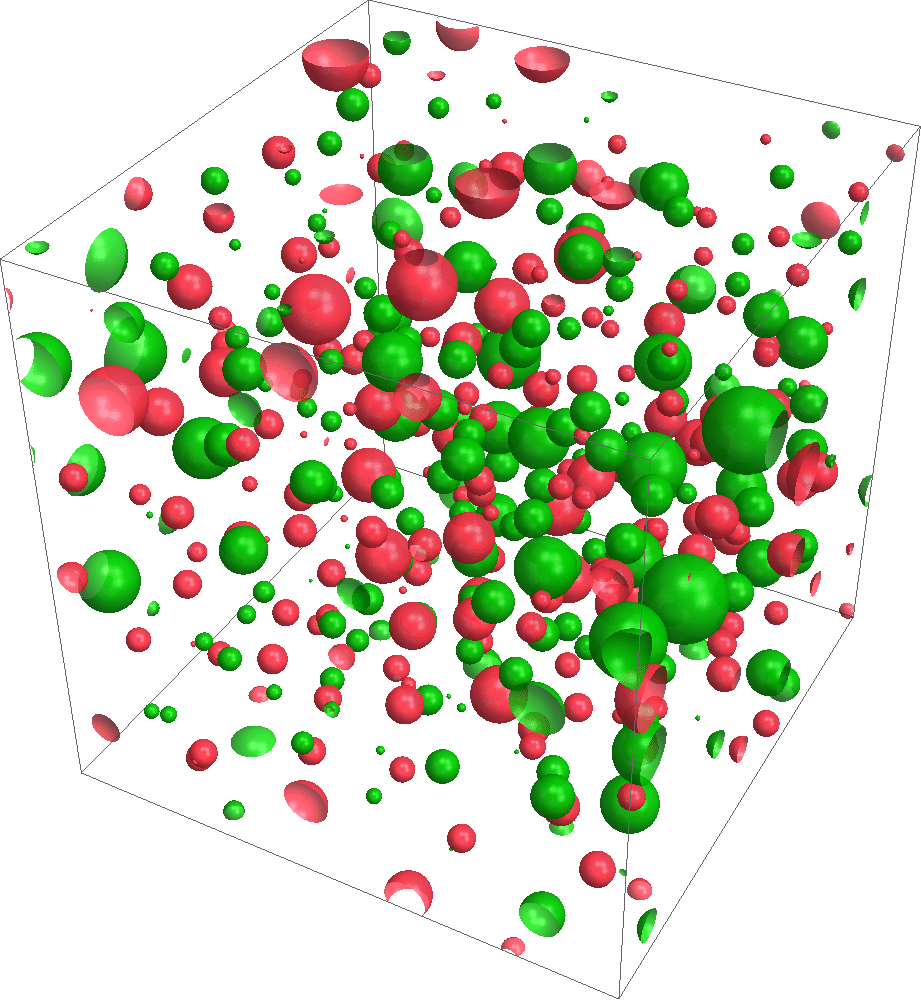}
			\label{het_ca_scm_low}
			}
		}
	\caption{
	Spatial distributions of the persistence times of the cation (Green) and the anion (Red) at the high temperature (T=6.24) (a) and the low temperature (T=1.12) (b). 
	The size of the sphere is proportional to the log of persistence time of each particle.
	The comparison of two snapshots clearly shows that persistence time is heterogeneous in time and also in space at the low temperature.
	}
	\label{het_ca_scm}
	\end{center}
\end{figure*}

In Fig.\ref{dis_tscan_scm}, the distribution of the cation in SCM is getting broader as the temperature is lowered.
Compared to the anion case, the cation clearly shows more heterogeneous dynamics at lower termperatures. 
(See the supporting information for the data of the ACM, UCM cases.)
The results 
is consistent  with the previous study that demonstrates the cation moves faster than the anion.\cite{park2015lifetime}
For the ACM case, the distribution of the anion is more heterogeneous than the cation.
This opposite result comes from the different charge distribution of the cation.
Compared to the SCM, the cations in the ACM make irregular structures around the anions because of their
asymmetric charge distribution.
As a result, it is expected that relatively small anion could have fast movement.
In the UCM system, alternating structure of the cations and the anions are not observed,
since there is no charge on the cations.\cite{park2015lifetime}
The cage effect would be suppressed and the distributions of two particles show the similar results.
The difference of two distributions are not profound 
but the anions have higher ratio of fast particles because of lower steric hindrance.

Fig.\ref{dis_scm} and Fig.\ref{g_s_cat} show the time dependence of the probability distributions
and the corresponding self-van Hove functions of the cations in SCM. 
A single peak at short time evolves into a broad distribution at $t\sim1\tau_{\alpha}$ and becomes a single peak again in the long time limit.
From this results, we can infer that the dynamics are most heterogeneous at time around from $t\sim1\tau_{\alpha}$ to $t\sim5\tau_{\alpha}$.
In the ACM and UCM, similar tendency is found for the calculated distributions.
At long time limit, the self-van Hove function is approaching the Gaussian distribution shown as a dashed line
in Fig.\ref{g_s_cat},
$G_s(\delta r;t)=(4\pi Dt)^{-3/2}\text{exp}(-\delta r^2/(4Dt))$, where $D$ is the diffusion coefficient.
However, even at $t=50\tau_{\alpha}$, there exists a mismatch at small and large $r$, which means that the dynamic heterogeneity still remains at this long time case.


\begin{figure*}[!htb]
	\begin{center}
	\mbox{
	\hspace{-20pt}
		\subfigure[]{
			\includegraphics[width=0.3\textwidth]{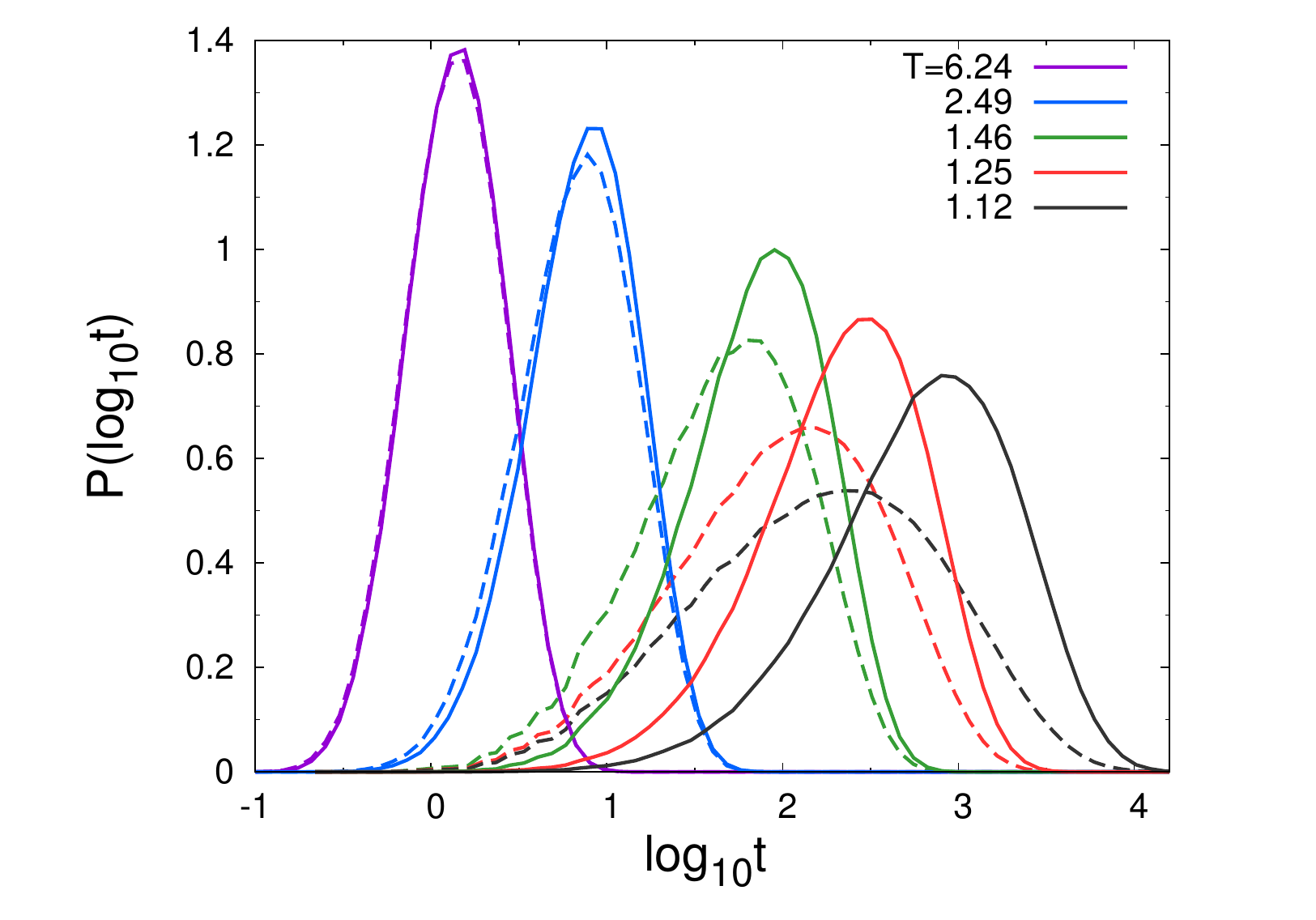}
			\label{per_exc_scm}
			}
	\hspace{-31pt}
		\subfigure[]{
			\includegraphics[width=0.3\textwidth]{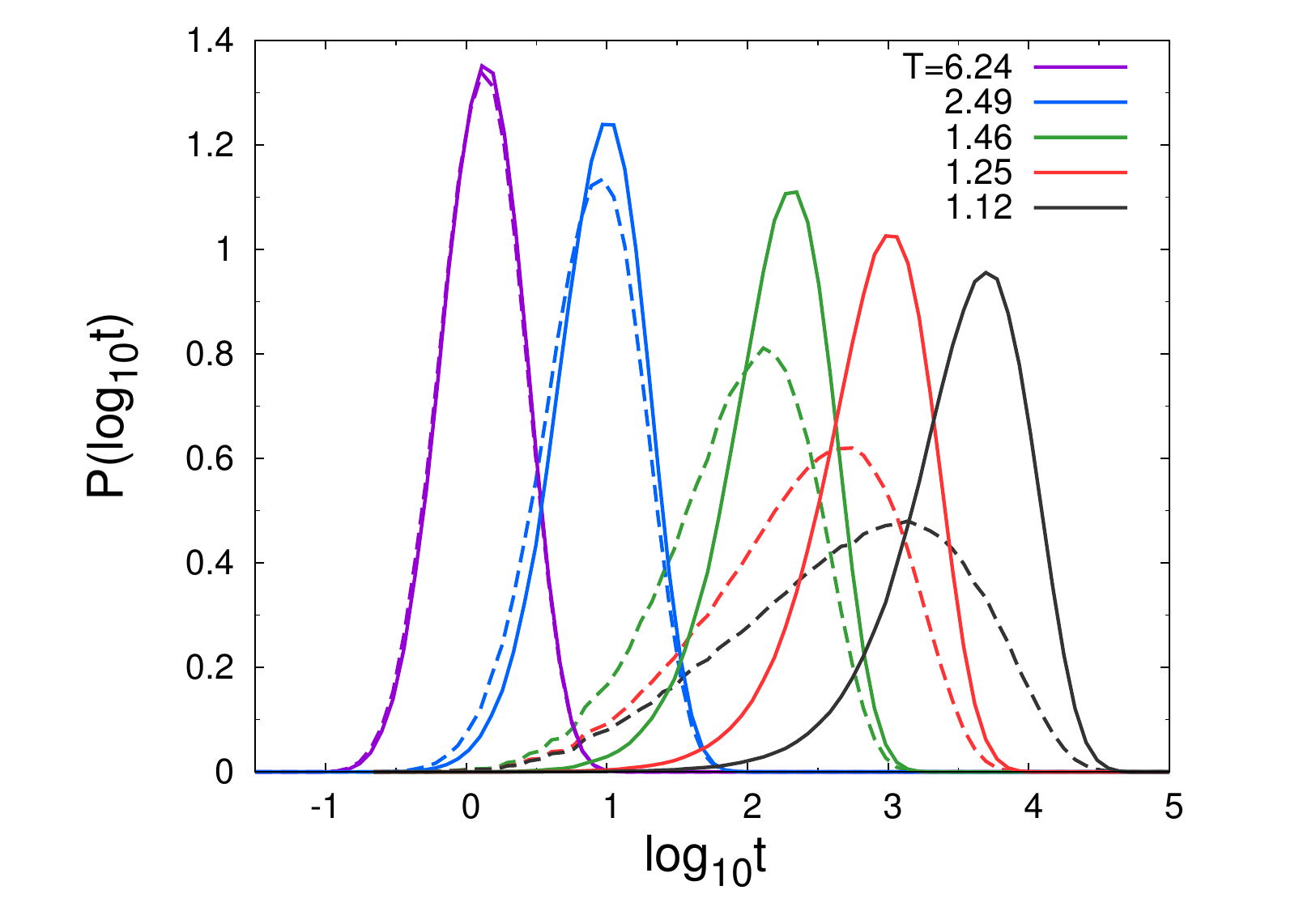}
			\label{per_exc_scm_ani}
			}
	\hspace{-29pt}			
		\subfigure[]{
			\includegraphics[width=0.3\textwidth]{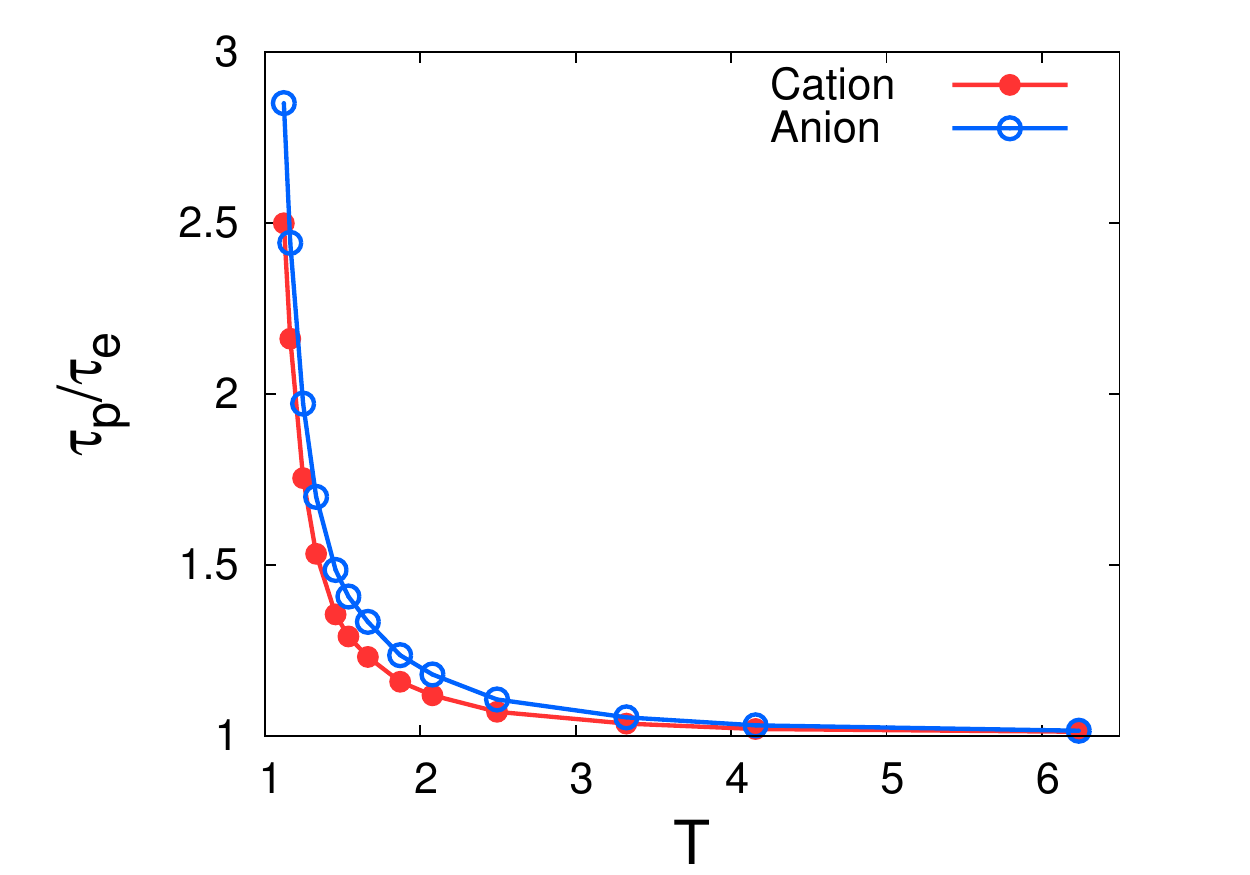}
			\label{violation}
			}
	\hspace{-40pt}	
	\vspace{10pt}	
					
		\subfigure[]{
			\includegraphics[width=0.31\textwidth]{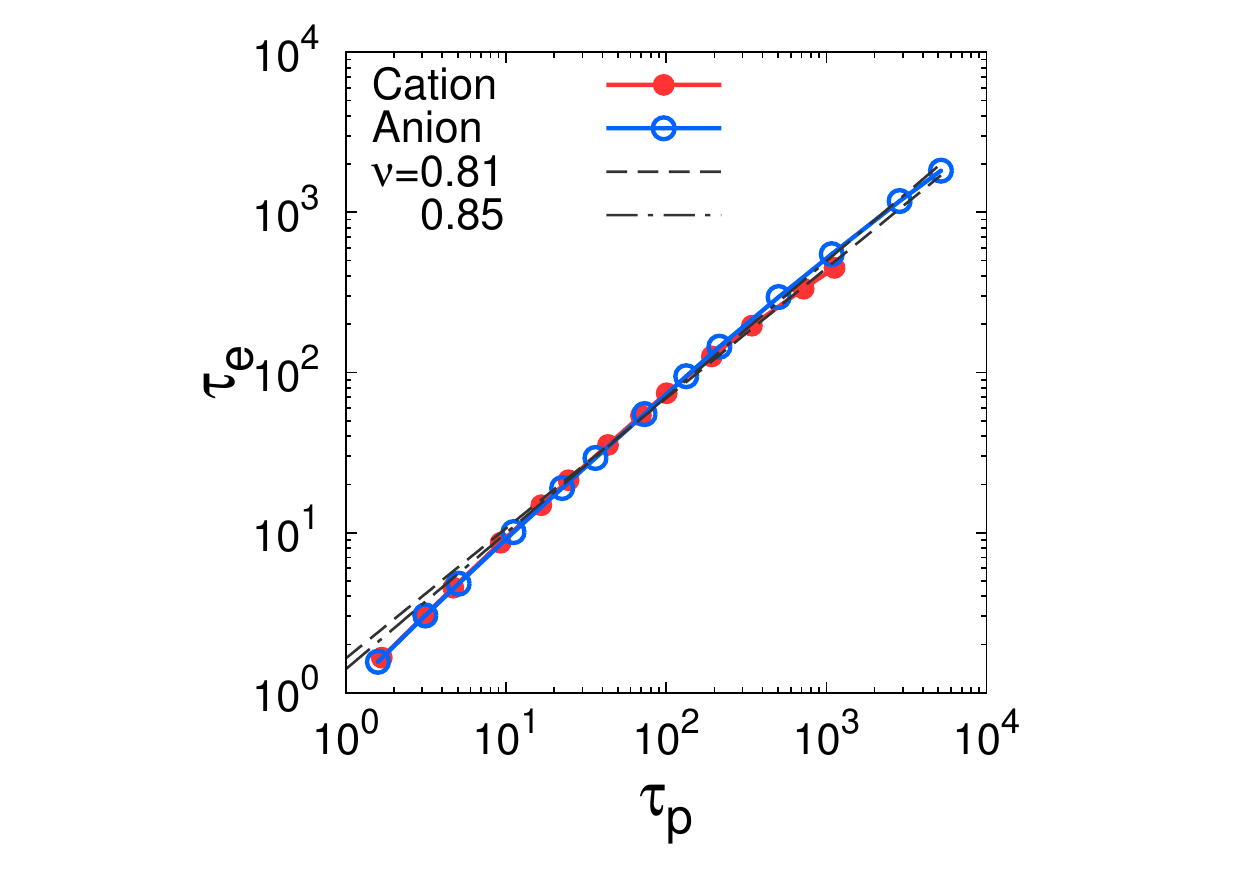}
			\label{per_exc_power}
			}
		}
	\caption{
	Probability distributions of the exchange time (dashed line) and the persistence time (solid line) for (a) the cation and (b) the anion in SCM.
	As the temperature is lowered, the distributions are decoupled.
	(c) The ratio of the mean persistence time and the mean exchange time increases abruptly at the low temperatures. (SCM)
	(d) The mean persistence time and the mean exchange time show power law relationship. (SCM) The data of ACM and UCM cases are presented in the supporting information.
	}
	\label{per_exc}	
	\end{center}
\end{figure*}

To characterize the dynamic heterogeneity in a different way, 
we calculate the excitation of each particle.\cite{jeong2010fragility,hedges2007decoupling}
The excitation is defined as an event that single particle $i$ moves more than distance $d$.
For example, when particle $i$ moved more than $d$ at time $t_1$,
$\lvert\mathbf{r}_i(t_1)-\mathbf{r}_i(0)\rvert>d$,
than the first excitation takes place at $t=t_1$.
Further, when particle moved more than $d$ from the $\mathbf{r}_i(t_1)$ after $t_2$,
$\lvert\mathbf{r}_i(t_1+t_2)-\mathbf{r}_i(t_1)\rvert>d$,
than the second excitation is at $t=t_1+t_2$.
For the third excitation at $t=t_1+t_2+t_3$, same rule is applied, 
$\lvert\mathbf{r}_i(t_1+t_2+t_3)-\mathbf{r}_i(t_1+t_2)\rvert>d$,
 and so on.
Using the series of the excitations, we can define two time scales, the persistence time and the exchange time.
The persistence time is the time value that the first excitation occurs,
so it is the set of all $t_1$ for every particle and trajectories.
Another time scale, the exchange time, is defined by the waiting time between two excitations.
It is the set of all $t_2, t_3, ...$.
It has been known that
the decoupling of the two time scales occurs when the system is dynamically heterogeneous.\cite{jeong2010fragility,hedges2007decoupling,jung2005dynamical}
Jung et al. first calculated the relationship of the two time scales in the kinetically constrained model (KCM)
and showed that the dynamic heterogeneity provokes the decoupling of the two time scales.\cite{jung2005dynamical}
Although the definition of the excitation is different from the case of the KCM,
the physical meanings are similar in the ionic liquids systems and it can be applied for studying the heterogeneous dynamics in those systems.
This analysis scheme has been applied to study the heterogeneous nature of the supercooled liquids, the ionic liquids, and the ring polymer melts.\cite{jung2005dynamical,hedges2007decoupling,jeong2010fragility,lee2015slowing}

Before analyzing the decoupling of the two time scales, we visualize the heterogeneous dynamics
using the persistence time of each particle.
Fig.\ref{het_ca_scm} shows the spatial distributions of the persistence times.
The radius of each particle represents the size of the persistence time in logarithm scale.
For the cufoff distance $d$, we used $d=1$ which is comparable to the size of the particles.
At the high temperature, Fig.\ref{het_ca_scm_high}, the persistence time distribution is 
relatively homogeneous than that at the low temperature, Fig.\ref{het_ca_scm_low}.
Note that not only the size of sphere is heterogeneous at the low temperature
but the spatial distribution also shows the heterogeneity.
This means that there is a correlation between the slow particles
and this could be the evidence of growing dynamic length scale.

Now, we investigate the decoupling of the two time scales and
relate this phenomenon with the breakdown of the Stokes-Einstein relation.
When the probability distribution of the exchange time is exponential,
the distribution of the two time scales is identical
because the persistence time is related with the exchange time through the integral.\cite{jung2005dynamical}
At high temperatures, there is a weak correlation between the excitation events,
which indicates that excitation events follow the Possion process.
At low temperature, however, the correlations between the excitation is pronounced
and the excitation events would experience non-Possion process
that results in decoupling of the persistence time and the exchange time distributions, Fig.\ref{per_exc_scm} and Fig.\ref{per_exc_scm_ani}.
We can interpret this phenomenon that the correlation between excitations increases because of the clustering of the slow particles.

The mean values of the two time scales also show the decoupling.
The mean persistence time and the mean exchange time, which are defined by the ensemble average of the persistence times, $\tau_{\text{p}}=\langle t_{\text{p}}\rangle$,
and the exchange times, $\tau_{\text{e}}=\langle t_{\text{e}}\rangle$,
are related to the transport coefficients, the relaxation time and the diffusion coefficient, respectively.\cite{jeong2010fragility}
It has been known that the mean persistence time would be proportional to the relaxation time, $\tau$,
when $d$ is comparable with $2\pi/q$ where $q$ is the first peak position of the structure factor.
Furthermore, the exchange event is governed by the diffusion of particle 
so that $1/\tau_{\text{e}}$ would have relation between diffusion constant, $D$, through power law relation.
Since the ionic liquid system is a kind of fragile liquids,\cite{park2015lifetime,jeong2010fragility}
it has been shown that there is a sublinear relation between $1/\tau_{\text{e}}$ and $D$, irrespective of $d$.\cite{jeong2010fragility}
This is due to the correlation between the excitation events.\cite{jung2005dynamical}
The power law relations between $\tau_{\text{p}}$ and $\tau_{\alpha}$,
and the relations between $\tau_{\text{e}}$ and $1/D$ are shown in supporting information.
Length scale $d$ dependence is also investigated in FIG.S7.
Fig.\ref{violation} shows the ratio of $\tau_{\text{p}}$ and $\tau_{\text{e}}$, 
which shows similar divergent behavior of $D\tau$ at low temperatures.\cite{park2015lifetime}
In addition, there are power law relations, $\tau_{\text{e}}\sim\tau_{\text{p}}^{\nu}$, between the two physical time scales
as it can be seen in Fig.\ref{per_exc_power}.
The value of the power law exponents are 
0.81 (SCM-cation), 0.85 (SCM-anion),
0.78 (ACM-cation), 0.77 (ACM-anion),
0.89 (UCM-cation) and 0.83 (UCM-anion).
The exponent of the SCM cation is analogous to that of the coarse-grained ionic liquids system
which is 0.80 for the cations.\cite{jeong2010fragility}
Comparing the values of the exponents,
we can infer that the fragility of the system increases in the order of UCM, SCM and ACM.
More detailed informations can be found in the supporting information.
\begin{figure*}[!htb]
	\begin{center}
	\mbox{
	\hspace{-40pt}
		\subfigure[]{
			\includegraphics[width=0.4\textwidth]{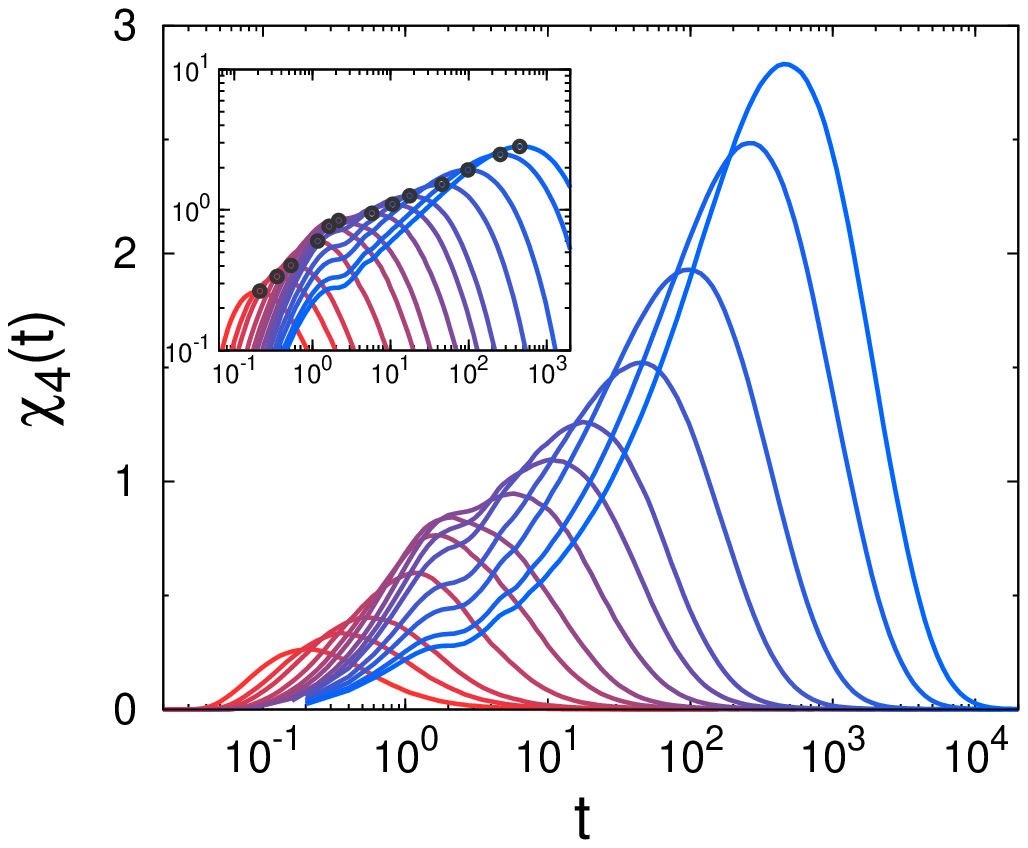}
			\label{chi4_scm_q}
			}
	\hspace{-36pt}
		\subfigure[]{
			\includegraphics[width=0.4\textwidth]{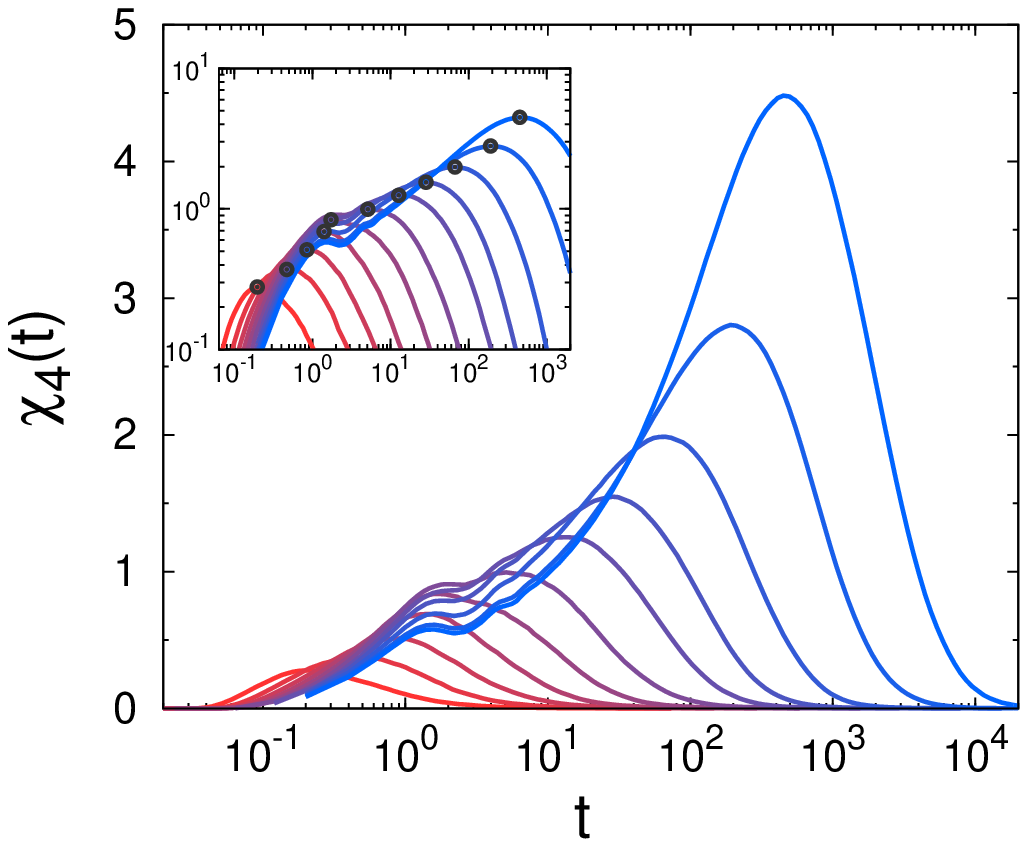}
			\label{chi4_acm_q}
		}
	\hspace{-32pt}
		\subfigure[]{
			\includegraphics[width=0.4\textwidth]{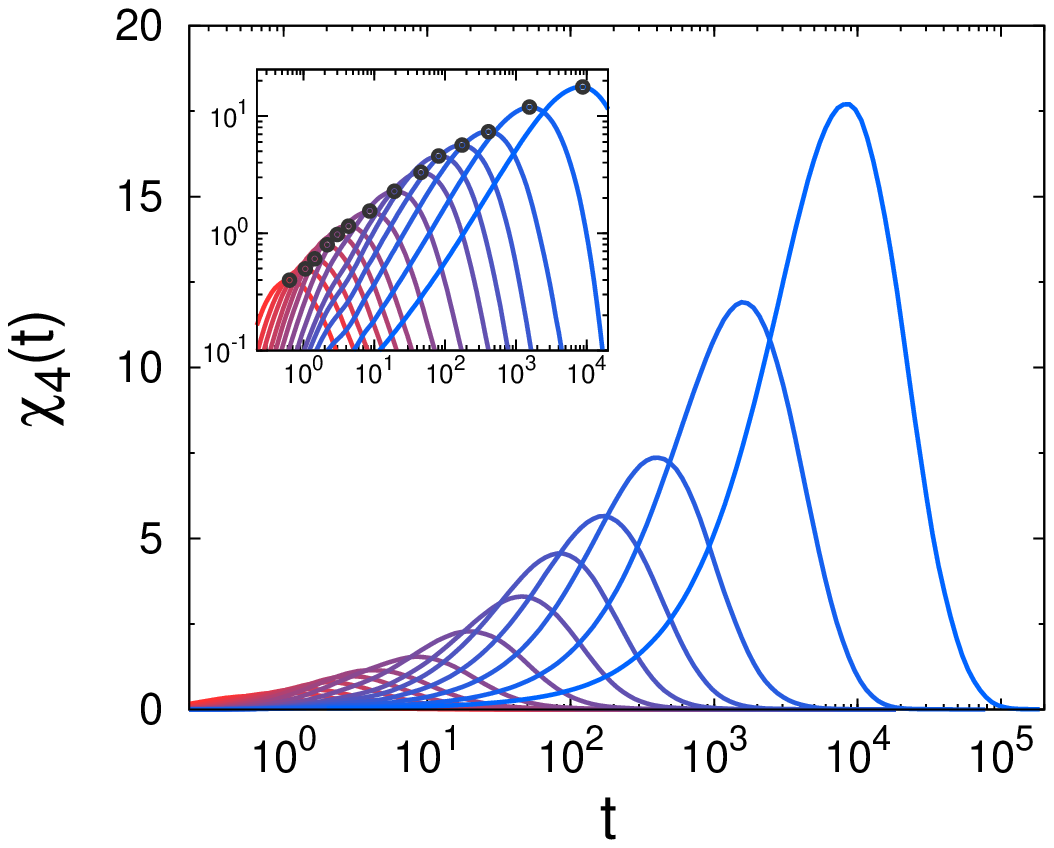}
			\label{chi4_ucm_q}
		}
		}
	\caption{
Dynamic susceptibility, $\chi_4(t)$, of the cations in each system calculated using Eq.\ref{eq-chi4} at various temperatures (a)SCM, (b)ACM and (c)UCM. The temperatures are the same with Fig.\ref{q_t}.
Log-log plots are also shown (inset).
The time value that makes $\chi_4(t)$ maximum is defined as the characteristic time scale of the dynamic heterogeneity, $t_4^*$ (black dots).
	}
	\label{dyn_sus}
	\end{center}
\end{figure*}

In this section,
we confirmed that the heterogeneous dynamics found in the supercooled liquids system are also found in 
our ionic liquids model systems.
From the displacement distribution, we find the clue that the mobility of the particles are heterogeneously distributed.
However, the correlations between slow or fast particles can not be obtained from this analysis.
In order to observe this correlated behavior, a four-point correlation function is introduced to calculate the dynamic length scale in the next section.
It is noteworthy that the distribution of the charge on the cations not only changes the local structure
but also affects the fragility of the whole system.
From the scaling analysis,
we find that the difference between the cation and the anion is smaller compared to the differences between the model systems.
Thus, we concentrate on the differences between the models rather than the type of ions.


\subsection{The dynamic susceptibility and the dynamic structure factor}

In the previous studies of glassy dynamics, 
there have been several different schemes to define the length scale of the dynamic heterogeneity.
\cite{donati1999spatial,donati2002theory,lavcevic2002growing,lacevic2003spatially,chandler2006lengthscale,flenner2011analysis,kim2013multiple}
Among them,
the dynamic length scale defined from the four-point correlation function has been widely adopted for many systems.
The four-point correlation function given in Eq.\ref{four-point} measures the correlation of the relaxation of the density fluctuation.
Since the dynamic susceptibility is obtained by integrating $g_4(\mathbf{r},t)$ over the space,
the dynamic susceptibility can be regarded as a volume of the correlated motion.
Furthermore, the dynamic susceptibility can be written in the form of the fluctuation of the dynamic quantity as will be shown below.
We employ this framework that has been previously established 
and applied to analyze the supercooled liquids systems.\cite{lacevic2003spatially,lavcevic2002growing,glotzer2000time}

\begin{figure}[!t]
	\begin{center}
	\mbox{
	\hspace{-30pt}
		\subfigure[]{
			\includegraphics[width=0.4\textwidth]{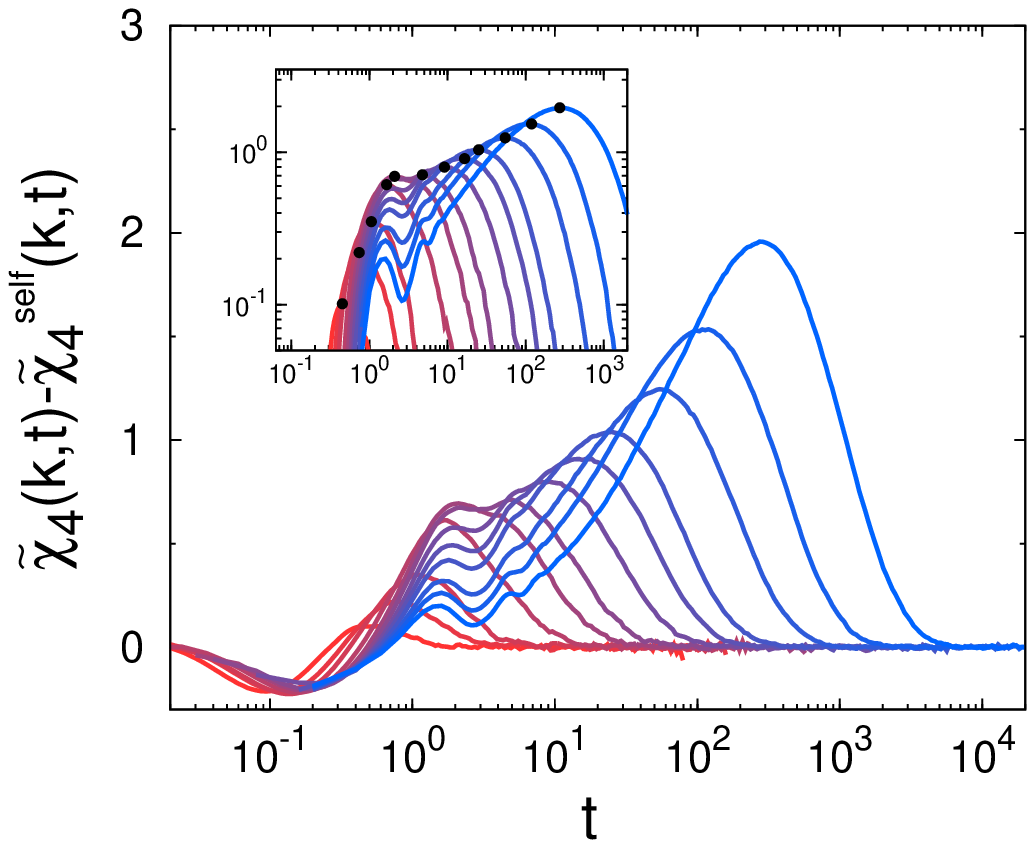}
			\label{dyn_sus_isf}
			}
	\hspace{-30pt}			
		\subfigure[]{
			\includegraphics[width=0.4\textwidth]{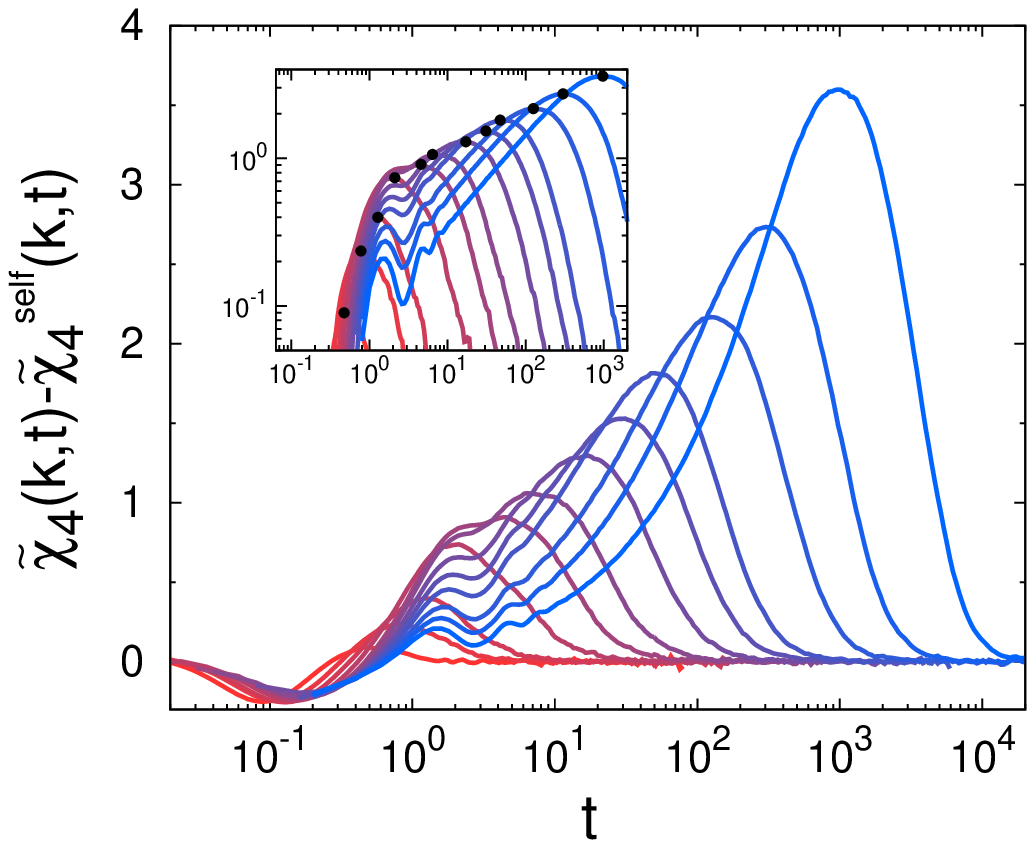}
			\label{dyn_sus_isf_ani}
			}			
		}
	\caption{
The dynamic susceptibility calculated using Eq.\ref{eq-chi4-isf} for (a) the cation and (b) the anion in SCM.
Wavevector $k$ is set by $k=2\pi/{\lambda}_{\text{max}}$,
where ${\lambda}_{\text{max}}=0.92$ is the shortest peak position of the radial distribution function between the cation and the anion.
The temperatures are the same with Fig.\ref{q_t}.
Log-log plot is illustrated in the inset and the maximum points are shown with black dots.
	}
	\label{dyn_sus_isfs}
	\end{center}
\end{figure}

\begin{figure*}[hbt]
	\begin{center}
	\mbox{
	\hspace{-40pt}
		\subfigure[]{
			\includegraphics[width=0.4\textwidth]{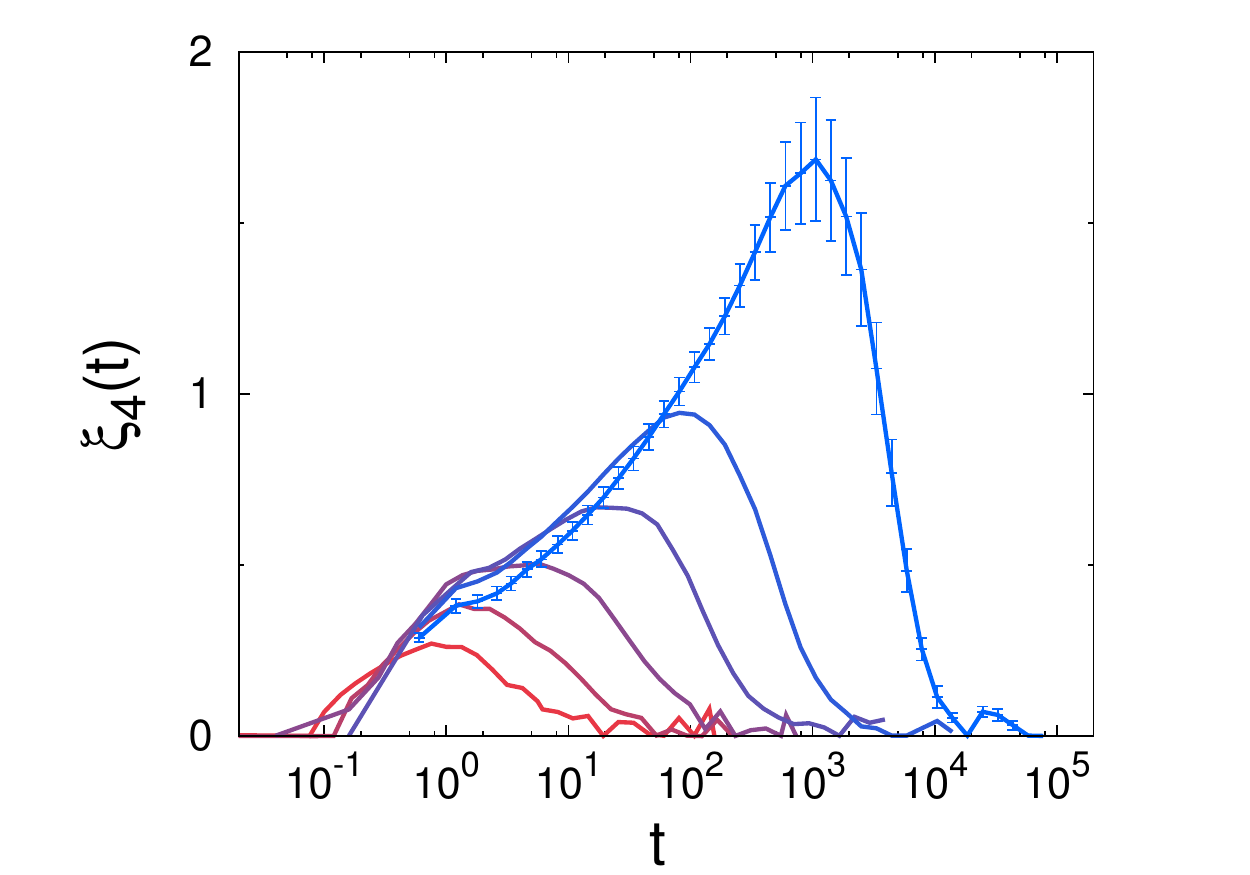}
			\label{}
			}
	\hspace{-36pt}			
		\subfigure[]{
			\includegraphics[width=0.4\textwidth]{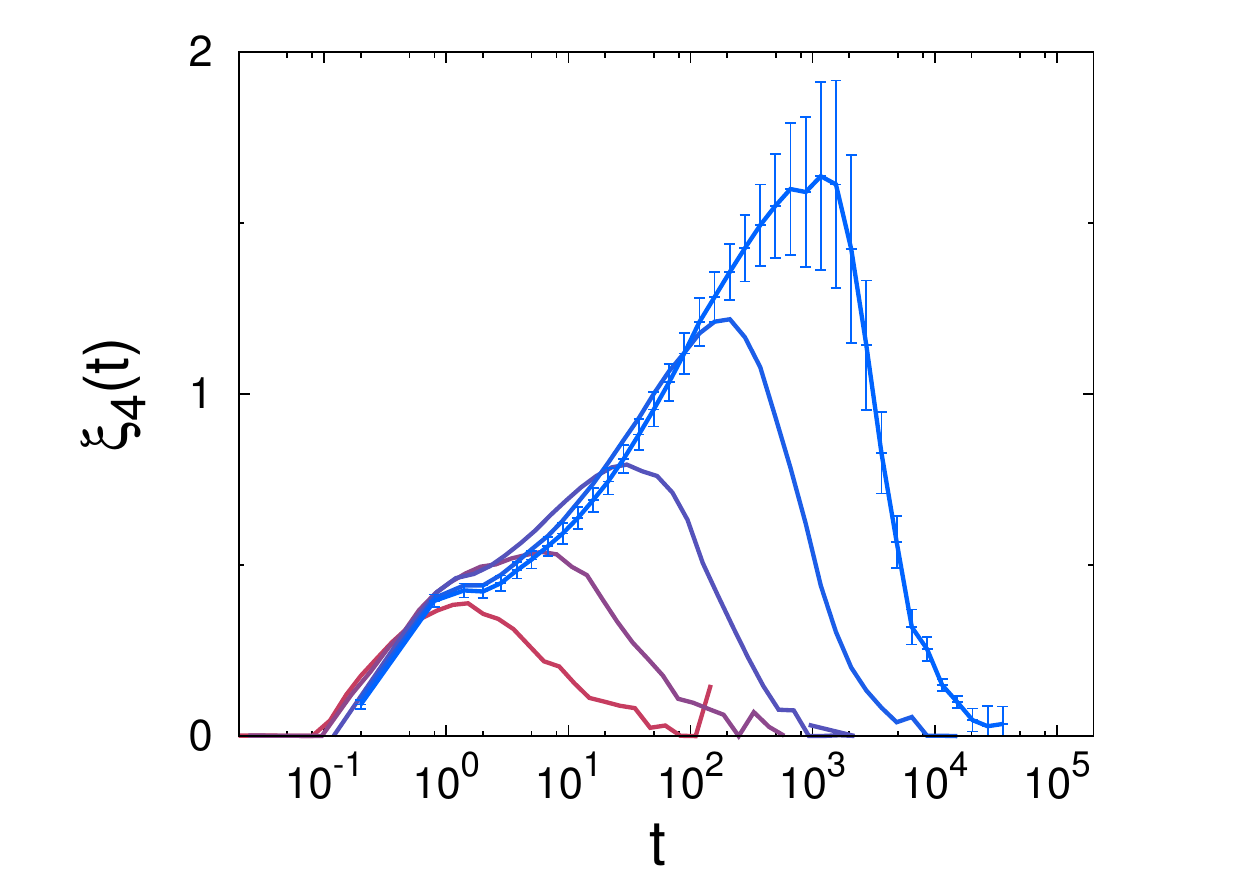}
			\label{}
		}
	\hspace{-32pt}		
		\subfigure[]{
			\includegraphics[width=0.4\textwidth]{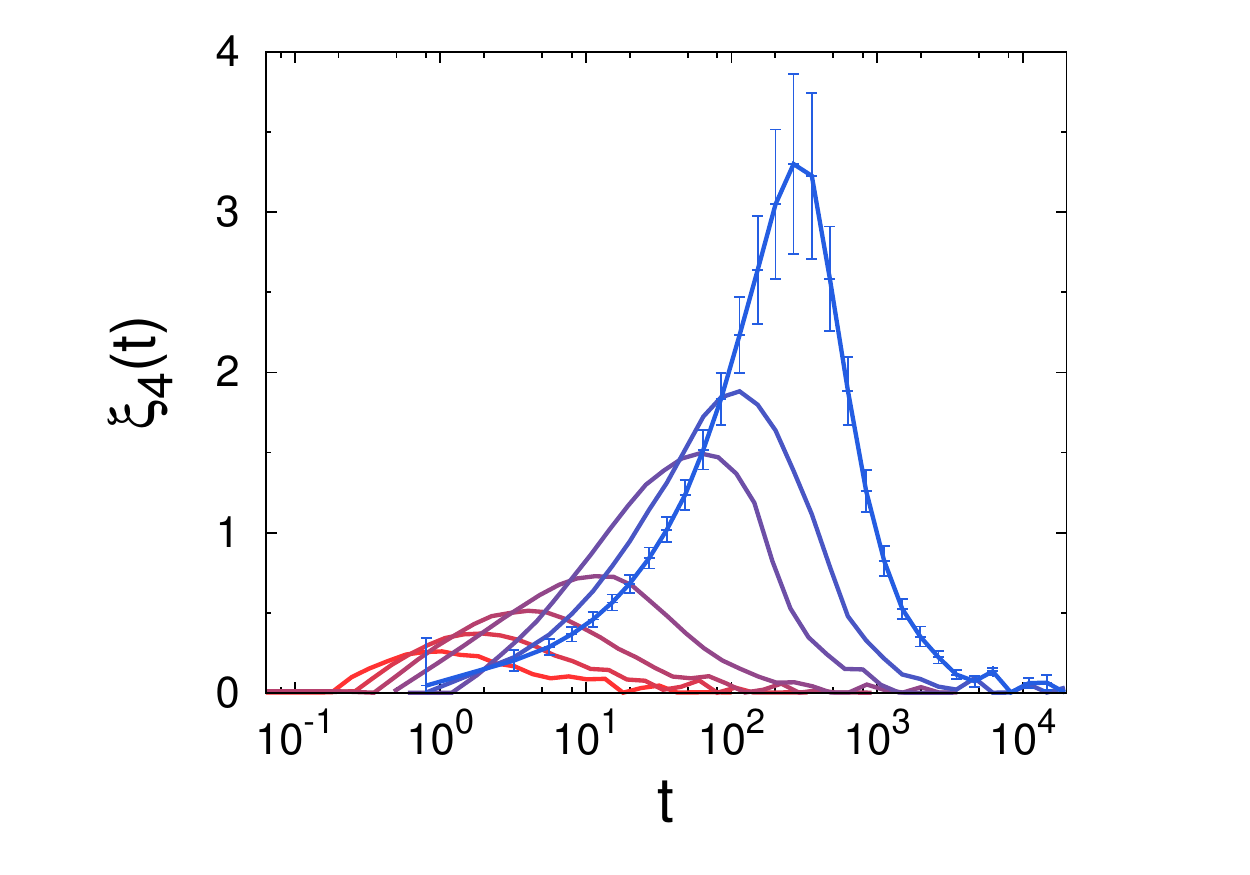}
			\label{}
		}
		}
	\caption{
Time dependence of the dynamic correlation length, $\xi_4(t)$, of the cation at various temperatures (a)SCM, (b)ACM and (c)UCM:
from left to right,
$T$= 6.24, 2.49, 1.87, 1.54, 1.33, and 1.16 (SCM),
$T$= 3.33, 2.49, 2.08, 1.87, 1.79, and 1.75 (ACM),
$T$= 1.56, 0.94, 0.67, 0.50, 0.40, 0.37, and 0.35 (UCM).
The error bars are shown only for the lowest temperature.
      	}
	\label{dyn_length}  	
	\end{center}
\end{figure*}

The dynamic susceptibility, $\chi_4(t)$, is defined as,
\begin{equation}
\label{eq-chi4}
\chi_4(t)=\frac{1}{N}\lbrack \langle Q(t)^2 \rangle- \langle Q(t) \rangle^2 \rbrack,
\end{equation}

 \begin{equation}
Q(t)=\sum_{i=1}^{N}w(\vert \mathbf{r}_i(0)-\mathbf{r}_i(t) \vert),
\label{eq-q_t}
\end{equation} 
where $w(\vert \mathbf{r}_i(0)-\mathbf{r}_i(t) \vert)$ is a overlap function
which is 0 when $\vert \mathbf{r}_i(0)-\mathbf{r}_i(t) \vert > a$ and
1 when $\vert \mathbf{r}_i(0)-\mathbf{r}_i(t) \vert \le a$.
$Q(t)$ counts the number of self overlapping particles using the configurations separated by a time interval $t$.
Thus, $Q(t)/N$ could be regarded as an index how much the system has been relaxed.
Fig.\ref{q_t} shows that $Q(t)/N$ decays from $1$ to $0$ as the time is passed, 
showing similar functional behavior with the self-intermediate scattering function.
In Fig.\ref{q_t}, we find that $Q(t)/N$ in charged systems shows more stretched form compared to that in the UCM. 
The difference between the model systems comes from the different local environment that the particles experience.
Detailed discussion will be given in Section 3.3.
Eq.\ref{eq-chi4} tells that $\chi_4(t)$ can be interpreted as a quantification of a fluctuation of $Q(t)$.
In this study, we choose overlap cutoff $a=0.3$ as used in other studies.\cite{glotzer2000time,lavcevic2002growing}
The $\alpha$-relaxation time, $\tau_{\alpha}$, can be defined as the time at which $Q(\tau_{\alpha})/N=1/e$.
This definition gives analogous value of $\tau_{\alpha}$ with the result of the conventional use of self-intermediate scattering function.
Additionally, for the cations, $\chi_4(t)$ calculated using the coordinates of
the center of mass and the particle itself did not show distinctive difference for the scaling law.
We will use the simulation data obtained by calculations with center of mass for each cation.

\begin{figure}[!ht]
	\begin{center}
	\mbox{
		\hspace{-20pt}
		\subfigure[]{
			\includegraphics[width=0.4\textwidth]{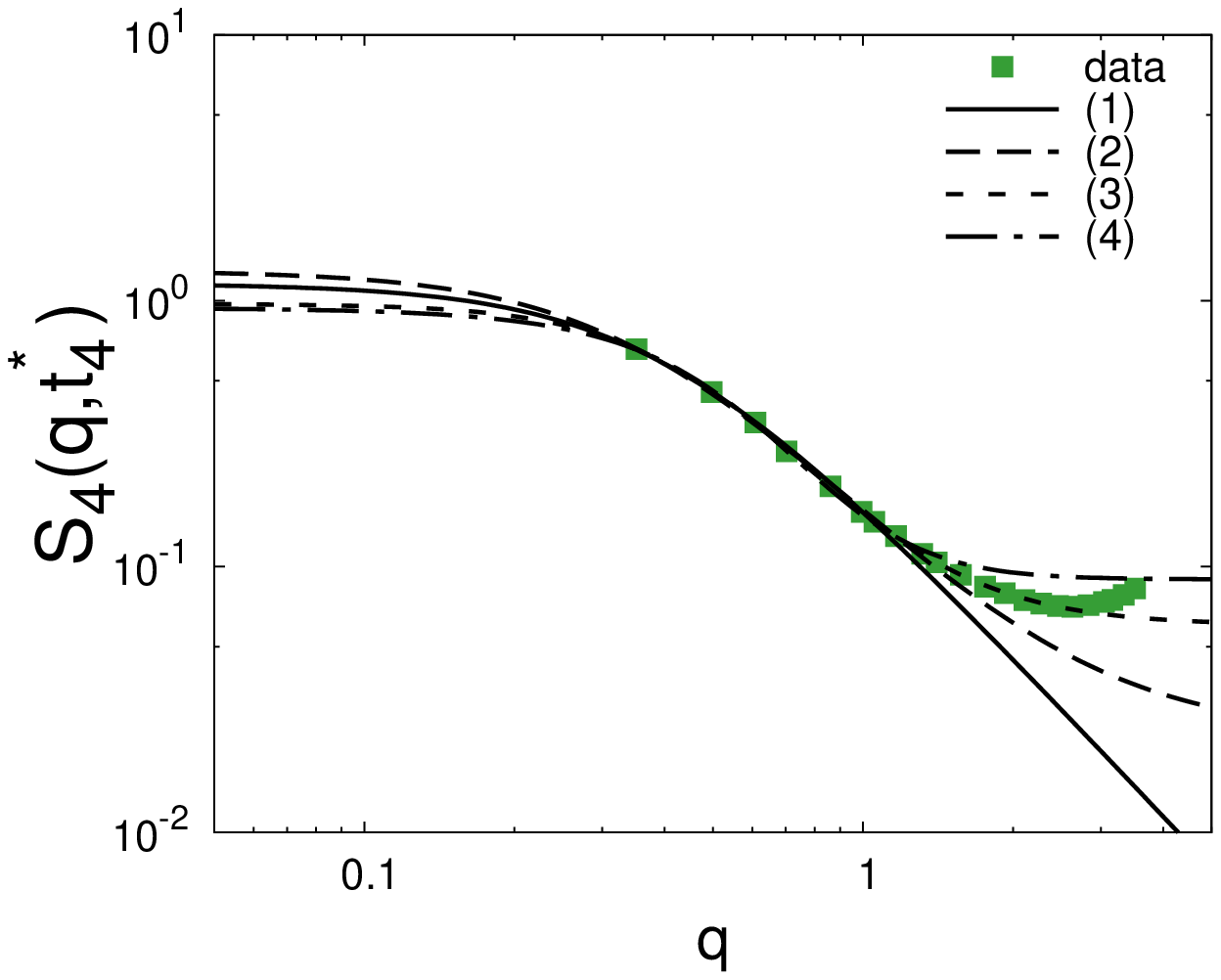}
			\label{str4_fit_0.56}
			}
		\hspace{-40pt}			
		\subfigure[]{
			\includegraphics[width=0.4\textwidth]{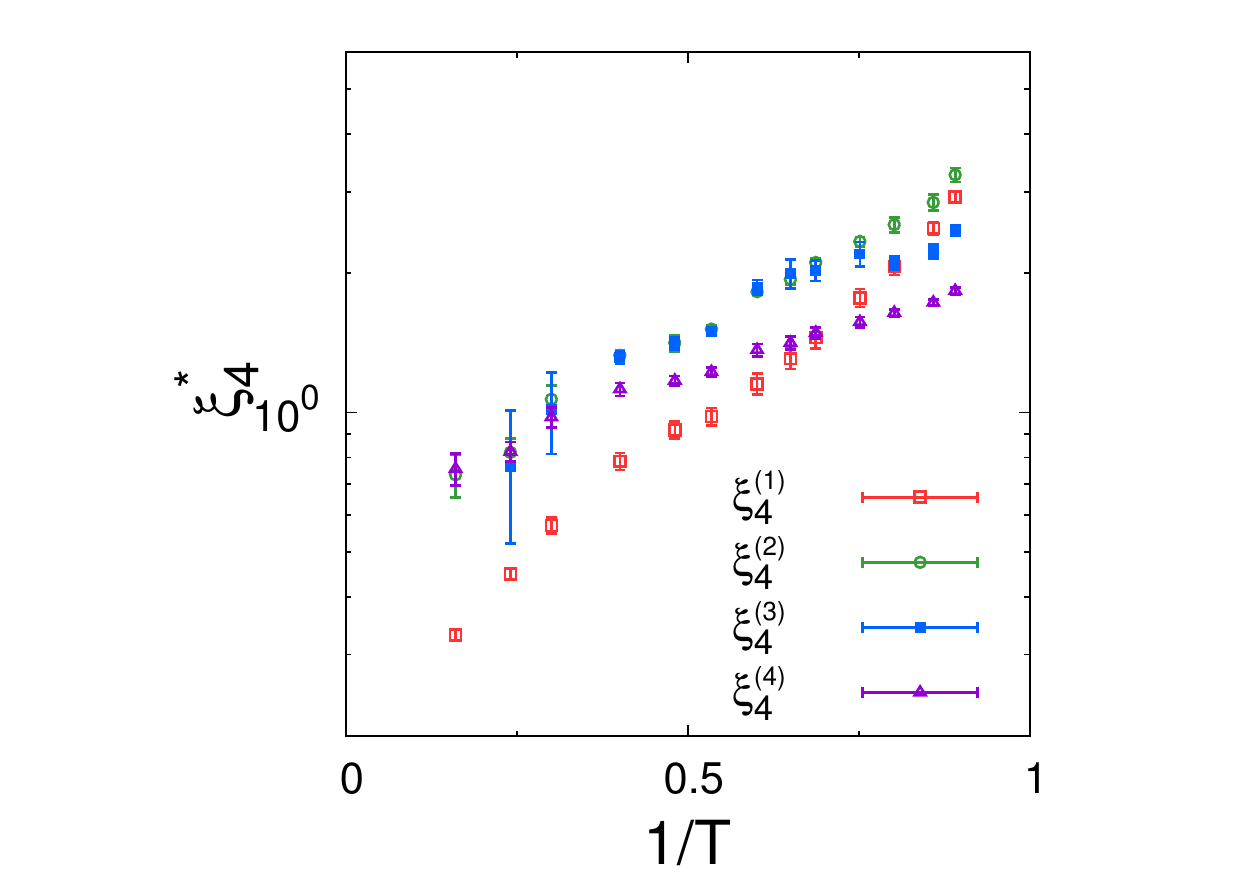}
			\label{dyn_length_compare_Temp}
			}
		}
	\caption{
Comparison of dynamic length scales obtained from different fitting schemes.
(a)Various fitting schemes on the data of the cation in SCM, at $T=1.16$.
(b)Temperature dependence of the dynamic correlation lengths.
      	}
	\label{dyn_length_compare}      	
	\end{center}
\end{figure}

Fig.\ref{dyn_sus} shows the dynamic susceptibility calculated using the Eq.\ref{eq-chi4} at various temperatures.
At a fixed temperature, $\chi_4(t)$ increases as the time passes. 
$\chi_4(t)$ has a maximum peak at a certain time scale, namely $t_{4}^{*}(T)$, which is comparable to the relaxation time, $\tau_{\alpha}(T)$.
We find that $t_4^*$ is proportional and almost equal to $\tau_{\alpha}$ for all the three model systems (See the supporting information).
This time scale, $t_4^*$, which shows the maximum value of the dynamic susceptibility, 
$\chi_4^*=\chi_4(t_4^*)$, is defined as a characteristic time scale of the dynamic heterogeneity.
As the time increases further, $\chi_4(t)$ decreases to zero.
The functional form of $\chi_4(t)$ confirms that the dynamic heterogeneity is transient in time.
As the y-axis in the inset of Fig.\ref{dyn_sus} is on log scale,
$\chi_4^*$ would show power law relation with $t_4^*$.
Note that the maximum values, $\chi_4^*$, are marked with black dots.
We find the power law relations between $t_4^*$ and $\chi_4^*$ for all systems.
Moreover, the crossover behaviors at short time regime are found for SCM and ACM.
The existence of crossover behavior is a unique phenomenon in our model compared to the models of supercooled liquids.
This phenomenon is based on the enhanced separation of sub-diffusive regime and diffusive regime for charged model.
The details about the crossover behavior will be discussed in the next section.

An alternative definition of the dynamic susceptibility can be used,\cite{chandler2006lengthscale}
\begin{equation}
\tilde{\chi}_4(k,t)=\frac{1}{N}\lbrack \langle \tilde{Q}(\mathbf{k},t)^2 \rangle- \langle \tilde{Q}(\mathbf{k},t) \rangle^2 \rbrack,
\label{eq-chi4-isf}
\end{equation}

\begin{equation}
\tilde{Q}(\mathbf{k},t)=\sum_{l=1}^{N}e^{i{\mathbf{k}}\cdot(\mathbf{r}_l(t)-\mathbf{r}_l(0))},
\label{eq-q-isf}
\end{equation}
where $k=\vert\mathbf{k}\vert$.
Here, $k$ is a wavevector that regulates the length scale of local area and has the similar role of $a$ in the overlap function.
Using the Eq.\ref{eq-chi4-isf} and Eq.\ref{eq-q-isf}, $\tilde{\chi}_4(k,t)$ can be expressed in the form, 
$(1/N) \sum_{j=1}^{N} \sum_{l=1}^{N} \langle {\delta}\tilde{Q}_{j}(\mathbf{k},t) {\delta}\tilde{Q}_{l}(-\mathbf{k},t) \rangle$,
where ${\delta}\tilde{Q}_{j}(\mathbf{k},t)=e^{i{\mathbf{k}}\cdot(\mathbf{r}_j(t)-\mathbf{r}_j(0))}-\langle e^{i{\mathbf{k}}\cdot(\mathbf{r}_j(t)-\mathbf{r}_j(0))}\rangle$.
In this form, the self part of $\tilde{\chi}_4(k,t)$ can be easily obtained applying $j=l$ condition, $\tilde{\chi}_4^{\text{self}}(k,t)=1-F_s(k,t)^2$, where $F_s(k,t)=\langle e^{i{\mathbf{k}}\cdot(\mathbf{r}_j(t)-\mathbf{r}_j(0))}\rangle$ is the self-intermediate scattering function.
Fig.\ref{dyn_sus_isfs} shows the $\tilde{\chi}_4(k,t)-\tilde{\chi}_4^{\text{self}}(k,t)$ for the cation and the anion in SCM at various temperatures.
The magnitude of the wavevector is set by $k=2{\pi}/{\lambda}_{\text{max}}$, where ${\lambda}_{\text{max}}$ is the shortest peak position of radial distribution function between the cation and the anion.
The overall functional form of $\tilde{\chi}_4(k,t)-\tilde{\chi}_4^{\text{self}}(k,t)$ is similar with $\chi_4(t)$, while the peak at short time scale is more pronounced.
The crossover behavior for the maximum point of $\tilde{\chi}_4(k,t)$ also can be found in the inset of Fig.\ref{dyn_sus_isfs}.


In order to obtain the length scale of the dynamic heterogeneity, we calculate the dynamic structure factor with the follwing equations,
\begin{equation}
S_4(q,t)=\frac{1}{N}\lbrack \rho(\mathbf{q},t)\rho(\mathbf{-q},t)\rbrack,
\label{eq-str4}
\end{equation}
 \begin{equation}
\rho(\mathbf{q},t)=\sum_{i=1}^{N}\text{exp}[i\mathbf{q}\cdot\mathbf{r}_{i}(0)]w(\vert \mathbf{r}_i(0)-\mathbf{r}_i(t) \vert),
\end{equation} 
where $q=\vert \mathbf{q} \vert$.
The dynamic correlation length, $\xi_4(t)$, is obtained by fitting the small wavevector regime 
into the Ornstein-Zernike (OZ) equation,
\begin{equation}
S_4(q,t)=\frac{S_4(q=0,t)}{1+(q\xi_4(t))^2},
\end{equation}
where, $S_4(0,t)$ and $\xi_4(t)$ are fitting parameters.
$S_4(q,t)$ is fitted in the regime of $q\le1.5$ which is corresponding to the condition, $4.19\le r\le17.88$.
Obtained correlation length is shown in Fig.\ref{dyn_length} at various temperature and time.
Similar to the dynamic heterogeneity, the correlation length is also transient in time.
At first, the correlation length is growing until it reaches the maximum,
and then it decreases.
The functional form of the correlation length resembles that of $\chi_4(t)$, however, 
the time values that the peaks occur are not identical.
Analogous to the previous studies on the supercooled liquids,
the time for which $\xi_4(t)$ is maximum is larger than $t_4^*$.\cite{toninelli2005dynamical,flenner2011analysis}
From this result, we find that the characteristic time scale of the correlation length is longer than the time scale of the dynamic heterogeneity.

\begin{figure*}[!ht]
	\begin{center}
	\mbox{
	\hspace{-40pt}	
		\subfigure[]{
			\includegraphics[width=0.4\textwidth]{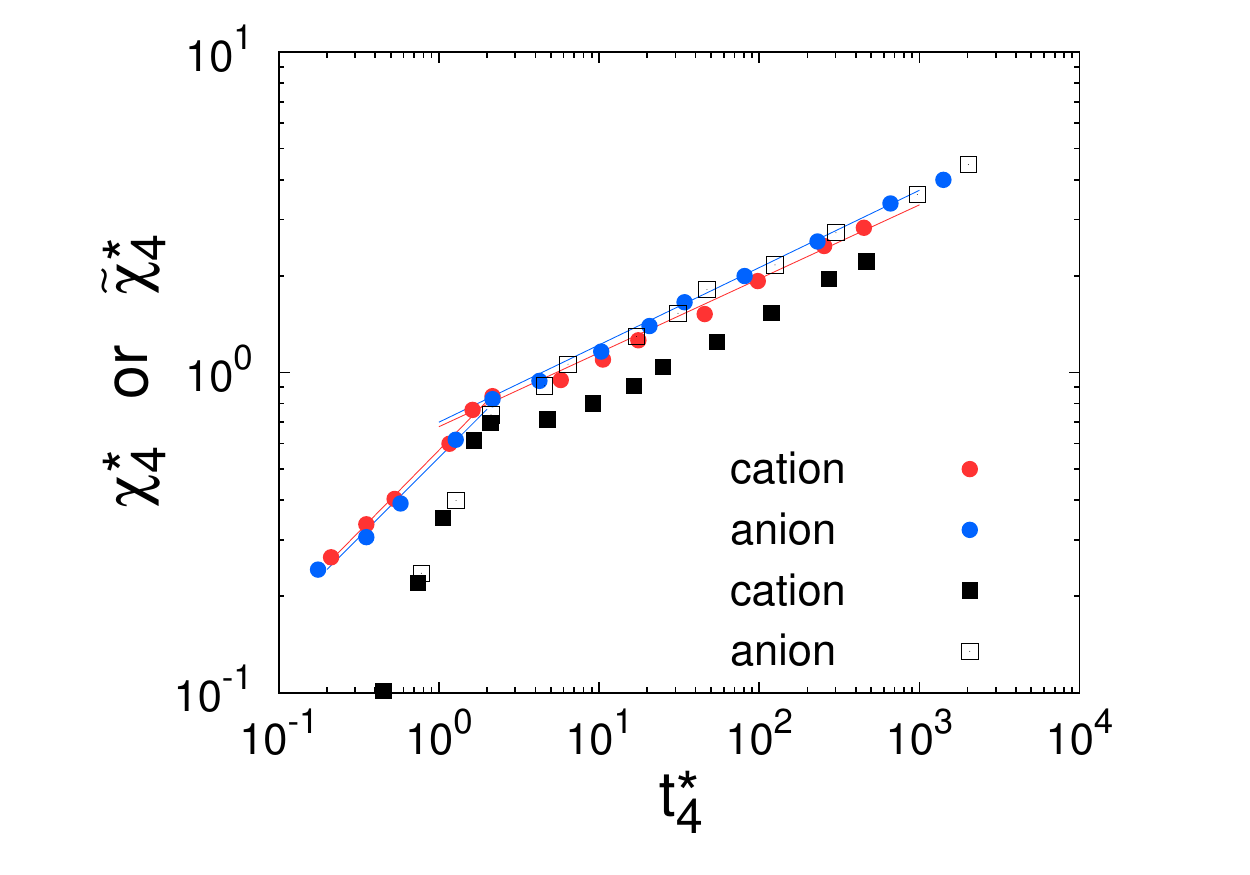}
			\label{}
			}
	\hspace{-36pt}				
		\subfigure[]{
			\includegraphics[width=0.4\textwidth]{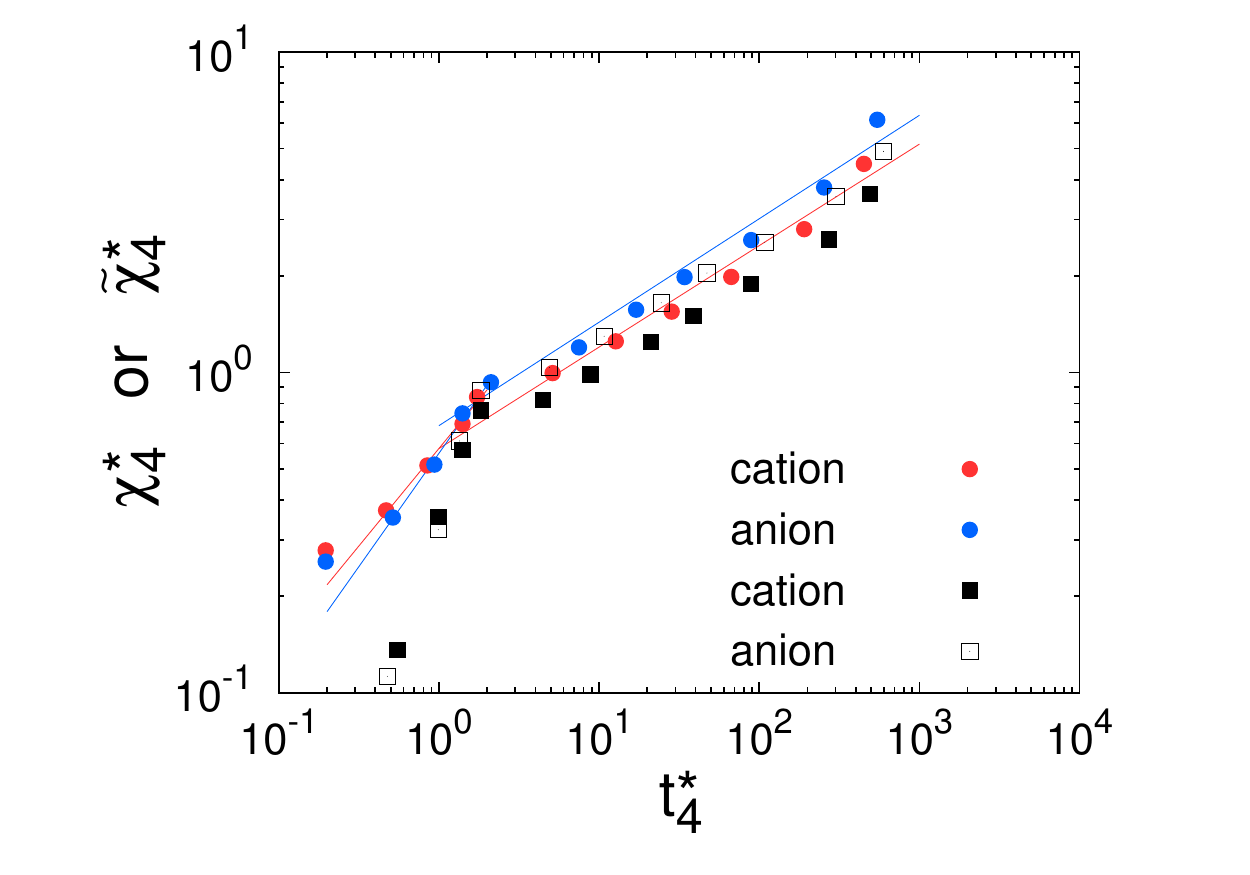}
			\label{}
		}
	\hspace{-32pt}			
		\subfigure[]{
			\includegraphics[width=0.4\textwidth]{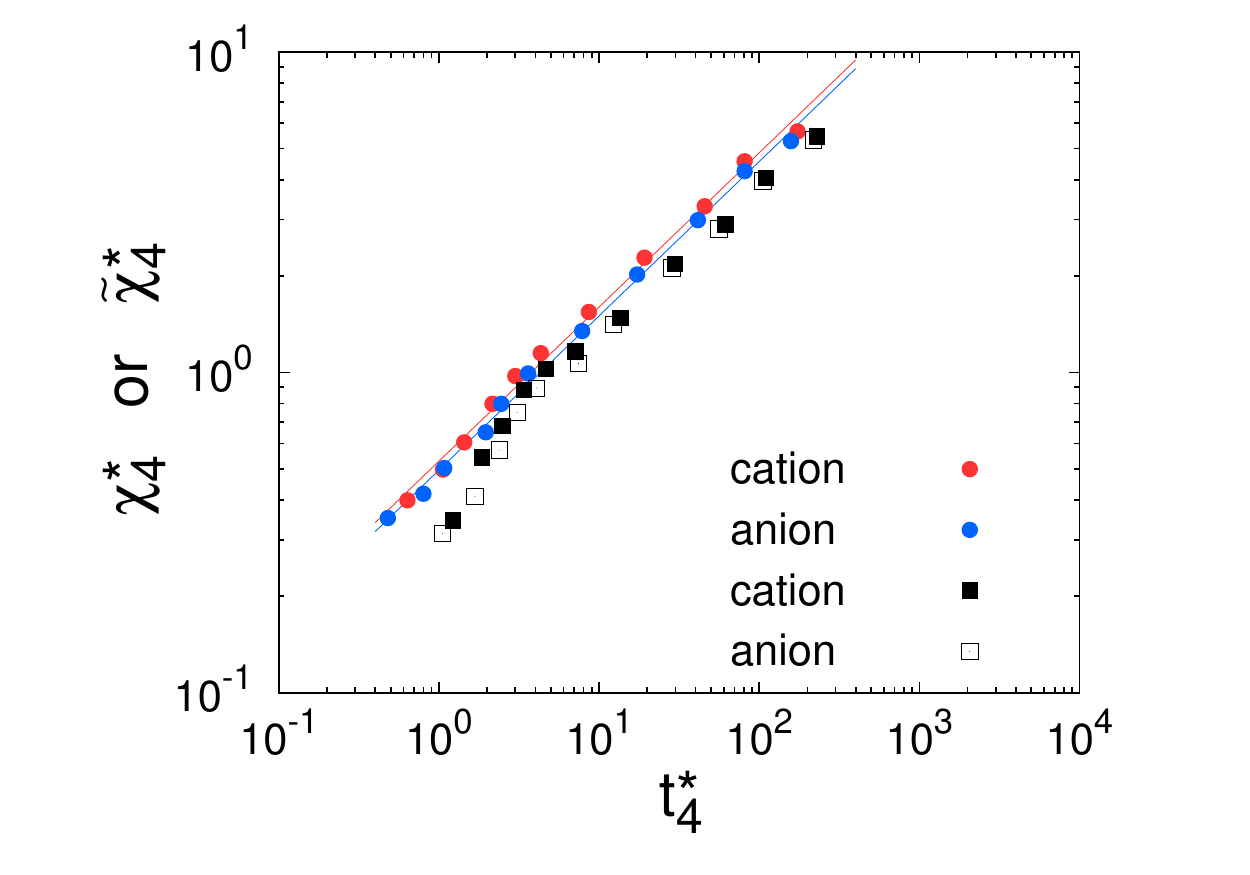}
			\label{}
		}
		}
	\caption{
Power law relation between $t_4^*$ and ${\chi}_{4}^*$ is shown for (a)SCM, (b)ACM and (c)UCM.
Circle denotes data using $\chi_4^*$ and square is for $\tilde{\chi}_4^*$.
Solid lines show power law fitting, $\chi_4^* \sim {t_4^*}^{\theta}$.
Power law exponents, $\theta$, are listed in Table \ref{table_exponent}.
The crossover behavior is profound in SCM and ACM.
The power law exponents using $\chi_4^*$ and $\tilde{\chi}_4^*$ are similar at the low temperature regime for all three models.
      	}
	\label{pow_t4_chi4}      	
	\end{center}
\end{figure*}

\begin{figure*}[!ht]
	\begin{center}
	\mbox{
	\hspace{-40pt}		
		\subfigure[]{
			\includegraphics[width=0.4\textwidth]{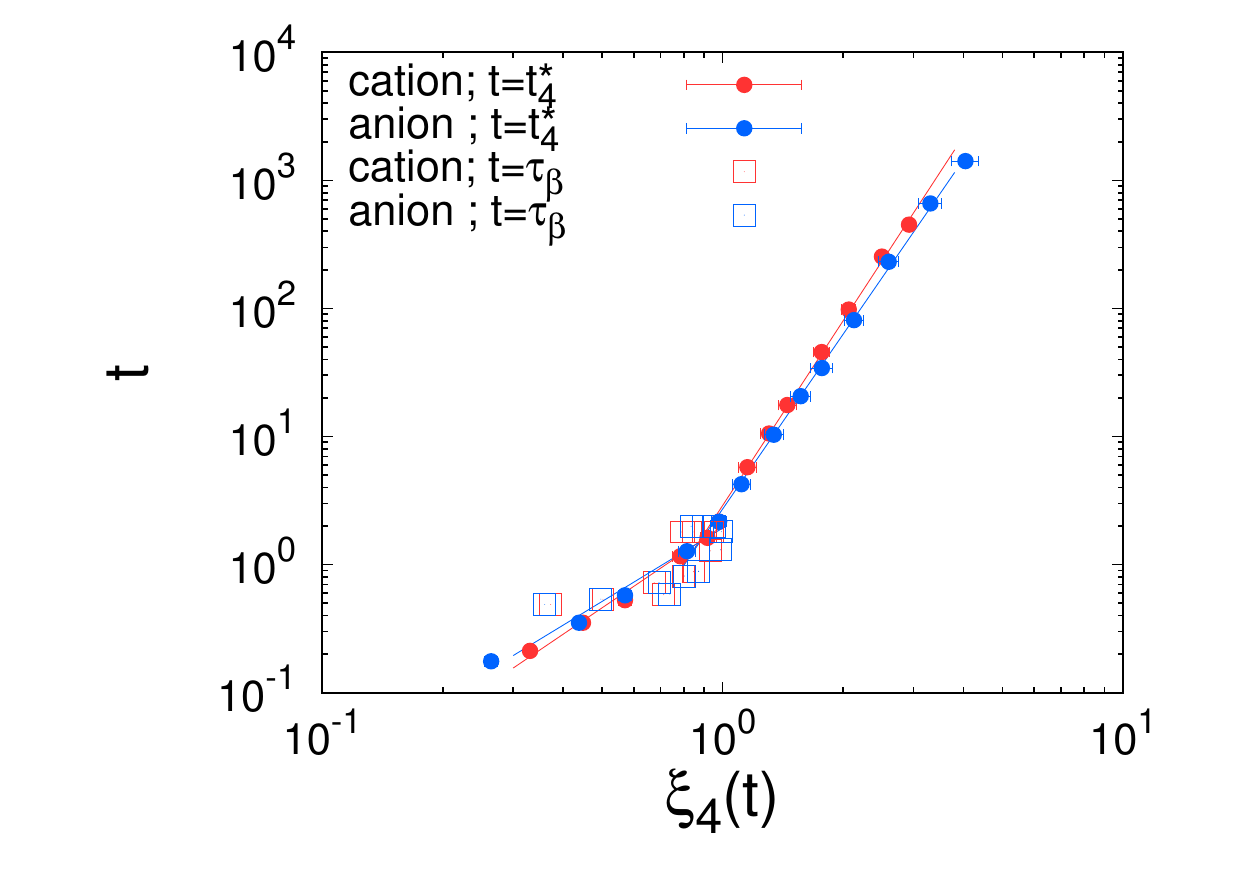}
			\label{}
			}
	\hspace{-36pt}					
		\subfigure[]{
			\includegraphics[width=0.4\textwidth]{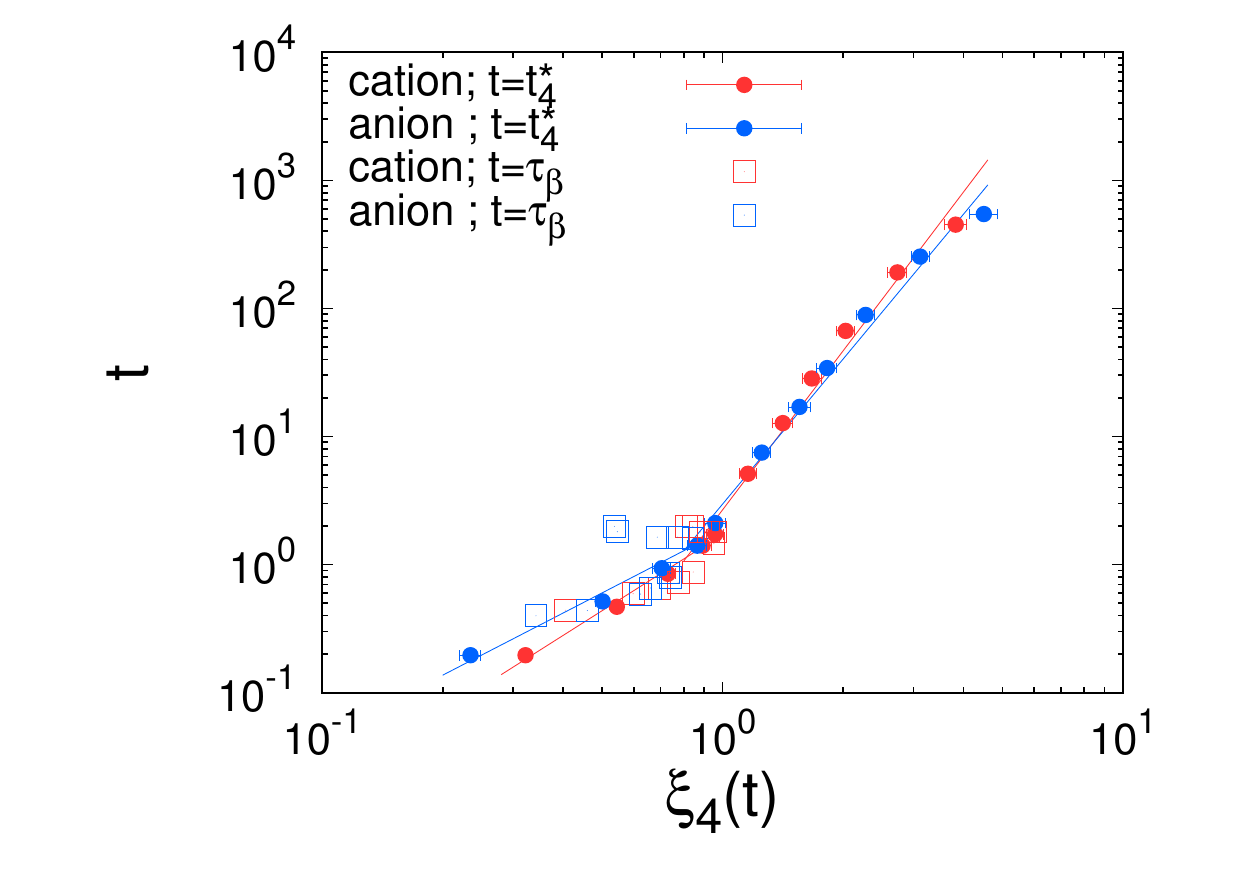}
			\label{}
		}
	\hspace{-32pt}				
		\subfigure[]{
			\includegraphics[width=0.4\textwidth]{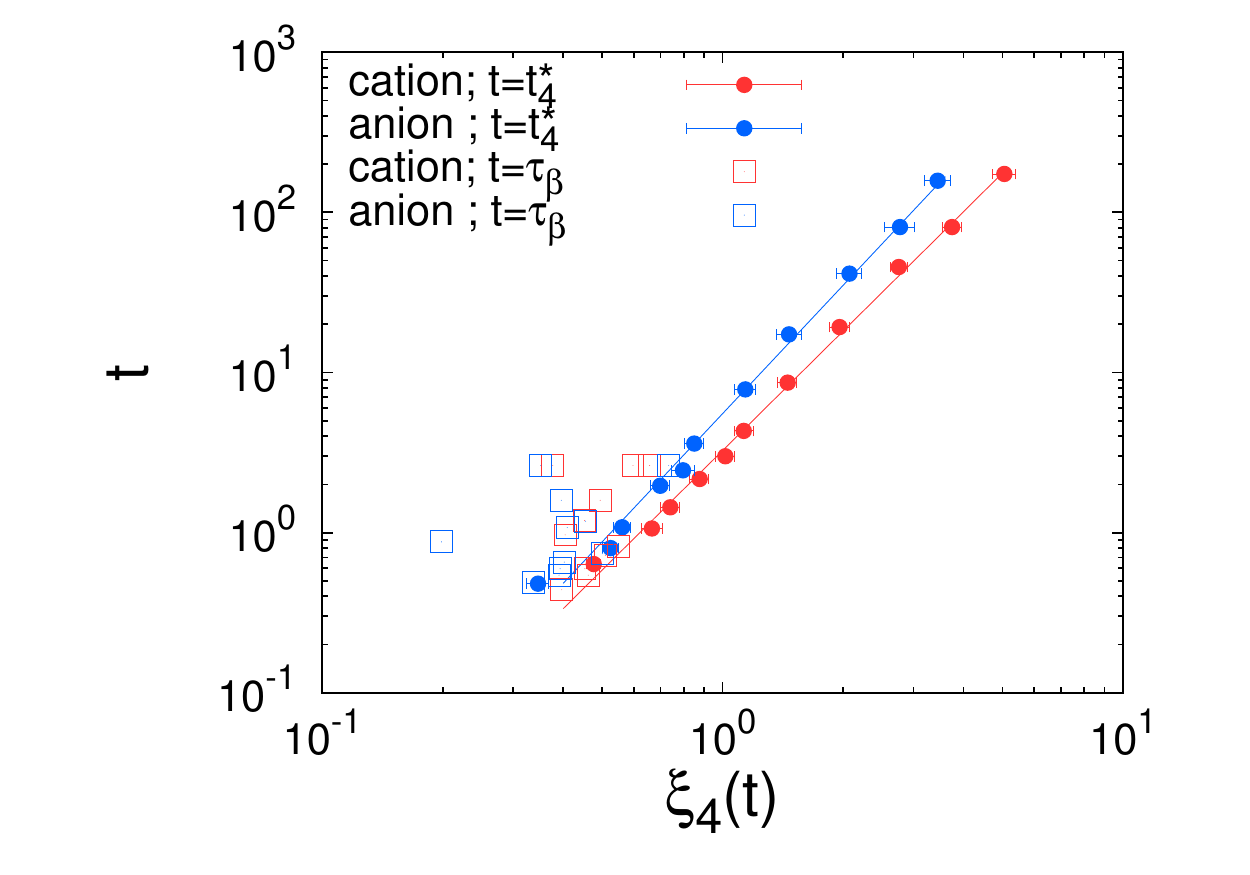}
			\label{}
		}
		}
	\caption{
Power law relation between $t_4^*$ and ${\xi}_{4}^*$ is shown for (a)SCM, (b)ACM and (c)UCM.
Circle denotes the data when $t=\tau_{\alpha}$ and square is for $t=\tau_{\beta}$.
Solid lines show power law fitting, $t_4^* \sim {\xi_4^*}^{\gamma}$.
Power law exponents, $\gamma$, are listed in Table \ref{table_exponent}.
The data at short length scale in condition of $t=\tau_{\alpha}$ (circle) correspond to the data in condition of $t=\tau_{\beta}$ (square) in SCM and ACM.
The crossover behavior is not observed in UCM.
      	}
	\label{pow_t4_xi4}      	
	\end{center}
\end{figure*}

In order to determine the dynamic correlation length of the system at a fixed temperature,
we use $\xi_4^*=\xi_4(t=t_4^*)$.
$\xi_4^*$ represents the dynamic correlation length when the dynamic heterogeneity is maximum.
All the dynamic structure factors collapse into single functional form of $f(x)=1/(1+x^2)$,
when the x-axis is $q\xi_4^*$ and y-axis is $S_4(q,t_4^*)/S_4(0,t_4^*)$. (See FIG.S11 in the supporting information)
The definition of the dynamic correlation length can be varied using different empirical functions.
We compare various correlation length obtained by fitting $S_4(q,t_4)$ into functions:
(1)$S_4(q,t_4^*)=S_4(0,t_4^*)/(1+(q\xi_4^{(1)})^2)$;
(2)$S_4(q,t_4^*)=(S_4(0,t_4^*)-C)/(1+(q\xi_4^{(2)})^2)+C$;
(3)$S_4(q,t_4^*)=(S_4(0,t_4^*)-C)/(1+(q\xi_4^{(3)})^{\zeta})+C$;
(4)$S_4(q,t_4^*)=(S_4(0,t_4^*)-C)/(1+(q\xi_4^{(4)})^2+(q\xi_4^{(4)})^4)+C$.
Here, the $C$ and $\zeta$ are fitting parameters and $C$ is included to improve the fitting against the baseline problem.\cite{toninelli2005dynamical} 
The fitted function to the data is shown in Fig.\ref{str4_fit_0.56} for the cation in SCM at T=1.16.

Fig.\ref{dyn_length_compare_Temp} shows the temperature dependence of these length scales for the cation in SCM.
Among these correlation lengths, $\xi_4^{(1)}$ grows faster than the other lengths
and shows clear crossover behavior.
Moreover, the scaling behavior of $\xi_4^{(1)}$ and $t_4^*$ reveals most reasonable power law exponent. 
In this sense, we use $\xi_4^{(1)}$ as a dynamic correlation length ,$\xi_4^*$ , in the rest part of this article.

\begin{figure}[!ht]
	\begin{center}
	\mbox{
	\hspace{-20pt}		
		\subfigure[]{
			\includegraphics[width=0.4\textwidth]{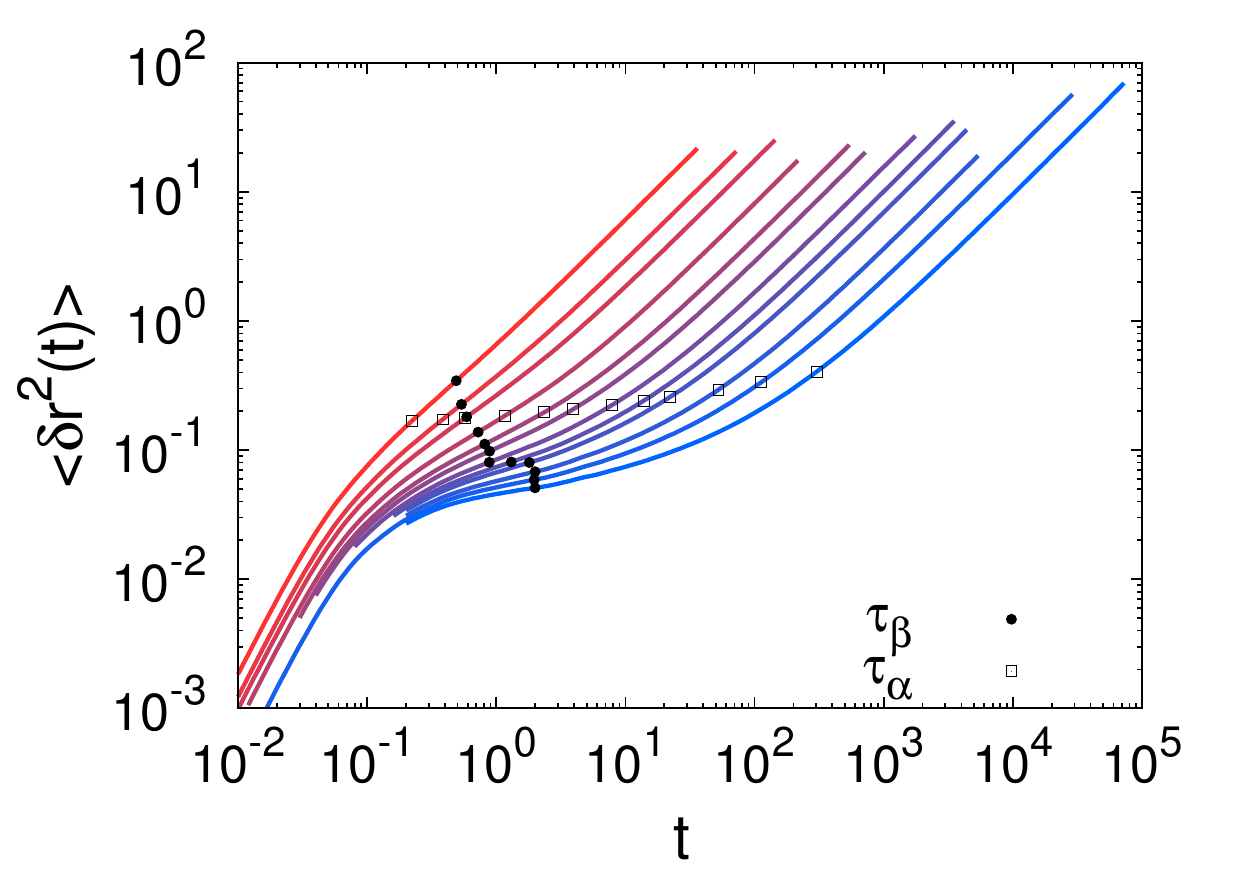}
			\label{msd_alpha_beta}
			}
	\hspace{-28pt}					
		\subfigure[]{
			\includegraphics[width=0.4\textwidth]{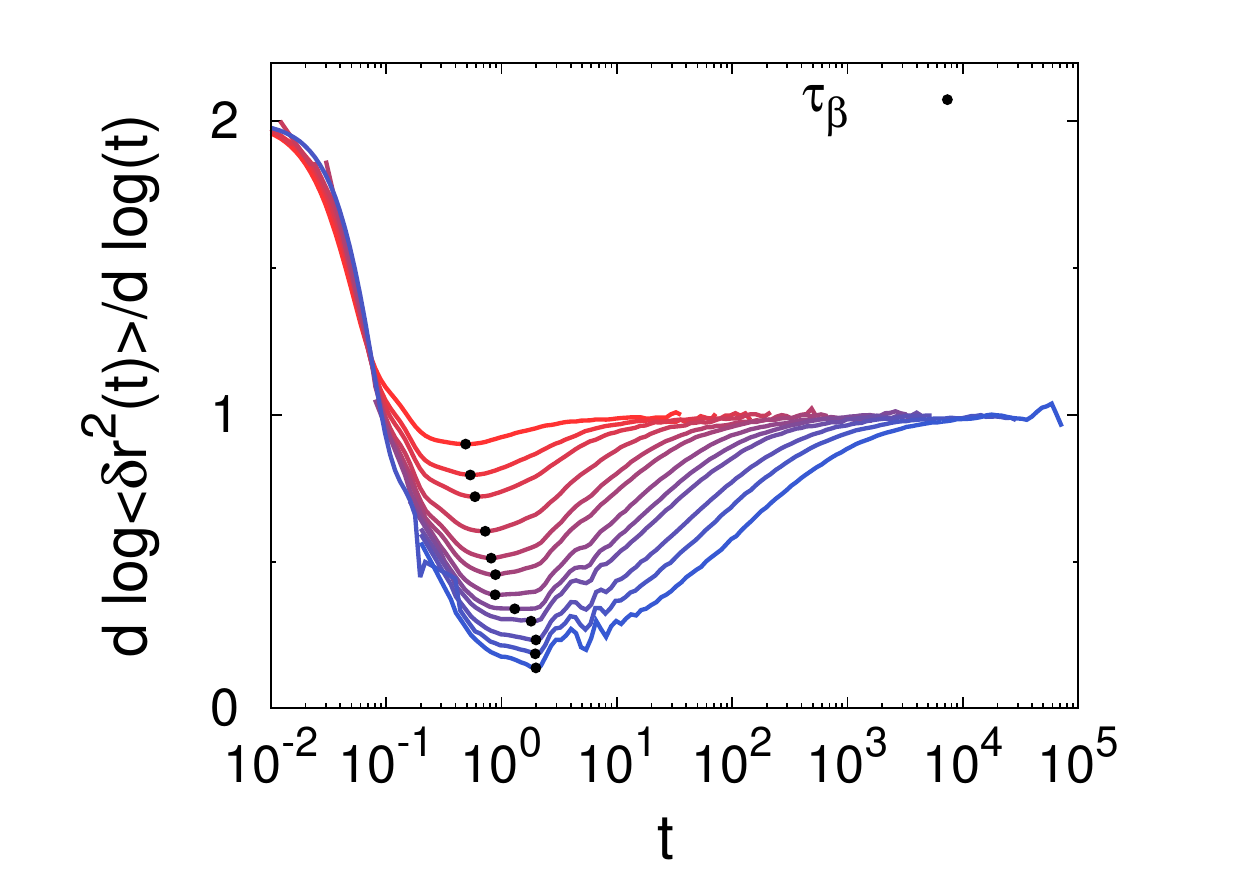}
			\label{msd_diff_alpha_beta}
			}
		}
	\caption{
Time scales ${\tau}_{\alpha}$ and ${\tau}_{\beta}$ are indicated with square and circle, respectively.
The cation in SCM at various temperature (same with Fig.\ref{chi4_scm_q}) is used.
(a)Mean-squared displacement $\langle \delta \mathbf{r}^2(t) \rangle$ is shown.
In the view of mean-squared displacement, $\tau_{\beta}$ is characteristic time scale for the plateau,
while $\tau_{\alpha}$ is the time scale of inset of diffusive regime.
(b)Derivative of mean-squared displacement is shown.
$\tau_{\beta}$ is defined as the time value that makes $\text{d}(\text{ln}\langle \delta \mathbf{r}^2(t) \rangle)/\text{d}(\text{ln} t)$ minimum.
      	}
	\label{cal_tau_beta}      	
	\end{center}
\end{figure}

\begin{table*}[!ht]
\center
  \caption{\ Various scaling exponents and corresponding equations}
  \label{table_exponent}
  \begin{tabular}{|c|c|c|c|c|c|c||c|}   
    \hline
    \multicolumn{2}{|c|}{} &\multicolumn{2}{c|}{SCM}&\multicolumn{2}{c|}{ACM}& \multirow{2}{*}{UCM} & \\ \cline{3-6}
    \multicolumn{2}{|c|}{} & high T & low T & high T & low T & & \\ \hline
    \multirow{2}{*}{\hspace{0.5cm}$\theta$\hspace{0.5cm}}&cation&0.51&0.24&0.61&0.33&0.48&\multirow{2}{*}{$\chi_4^* \sim {t_4^*}^{\theta}$} \\
    &anion& 0.50& 0.25&0.71&0.33&0.48&  \\ \hline
    \multirow{2}{*}{$\gamma$}&cation&2.1&4.8&2.0&4.1&2.5& \multirow{2}{*}{$t_4^* \sim {\xi_4^*}^{\gamma}$} \\
    &anion&1.9&4.5&1.6&3.8&2.7&  \\ \hline 
  \end{tabular}

\end{table*}

\subsection{Scaling laws}
We now investigate the power law relations between the dynamic physical quantities we calculated.
In various model systems of the supercooled liquids, 
the scaling law has been found for the dynamic length and time scales.\cite{stein2008scaling,whitelam2005renormalization,berthier2005numerical,whitelam2004dynamic,biroli2006inhomogeneous}
It is noted that these kinds of relations are originally found in the critical behavior of phase transitions.
In our ionic liquid model system, a similar power law relation is discovered.
First, we show the relation between $t_4^*$ and $\chi_4^*$ in Fig.\ref{pow_t4_chi4}.
For all three systems, the power law relation, $\chi_4^* \sim {t_4^*}^{\theta}$ is found.
Another power relation for $t_4^*$ and $\xi_4^*$ is also found, $t_4^* \sim {\xi_4^*}^{\gamma}$, as shown in Fig.\ref{pow_t4_xi4}.
The power law exponents are listed in Table \ref{table_exponent}.
As previously observed, there is clear crossover behavior.
Interestingly, this crossover behavior is prominent in SCM and ACM which include the charge on the particles.
Kim et al. also found the crossover behavior for a glass-forming binary soft-sphere mixture,
and addressed that this is due to different physical behaviors at different time scales, which are $\beta$-relaxation time, $\tau_{\beta}$, and $\alpha$-relaxation time, $\tau_{\alpha}$.\cite{kim2013dynamic}
In the article, $\tau_{\beta}$ is defined by the minimum value of $\text{d}(\text{ln}\langle \delta \mathbf{r}^2(t) \rangle)/\text{d}(\text{ln} t)$, where $\delta \mathbf{r} = \mathbf{r}(t)-\mathbf{r}(0)$.
This means that $\tau_{\beta}$ is the time value of the plateau of mean-squared displacement at each temperature.
The mean-squared displacement and its derivative of the cation in SCM is shown in Fig.\ref{cal_tau_beta}.
As the temperature is lowered, $\tau_{\beta}$ increases until it reaches maximum value of $t=2$.
Note that our definition of $\tau_{\alpha}$ and $\tau_{\beta}$ allows $\tau_{\beta}$ to be longer than $\tau_{\alpha}$ at the high temperatures.
$\tau_{\alpha}$ and $\tau_{\beta}$ are also marked in $Q(t)/N$, Fig.\ref{q_t}.
Note that $\tau_{\alpha}$ is a time scale of the structural relaxation, while $\tau_{\beta}$ is a time scale of plateau of $Q(t)/N$. 

In this sense, $\tau_{\beta}$ can be interpreted as a characteristic time scale that particles stay in the cage.
When the temperature is high, ${\beta}$-relaxation regime is not clearly observed in the time correlation function $Q(t)/N$.
However, at the low temperature, slowing down of local dynamics due to the cage effect makes
${\beta}$-relaxation regime distinctive.
The onset temperature of this phenomenon is related to the onset of the crossover behavior.
When $\xi_4(t)$ is calculated at $t={\tau}_{\beta}$, 
the cations and the anions of SCM and ACM have the power law exponent ${\gamma} \sim 2$ (Fig.\ref{pow_t4_xi4}).
However, when $t=t_4^*\sim\tau_{\alpha}$, the exponent is much larger at low temperatures.
Such phenomenon is not found in the UCM system since the cage effect of uncharged system is weak compared to the charged systems.
The evidence of enhanced cage effect in SCM and ACM can be found in the behavior of $Q(t)/N$, Fig.\ref{q_t}.
For SCM and ACM, $Q(t)/N$ show highly stretched form and the plateau is clearly observed at short time.
However, the plateau in $Q(t)/N$ of UCM is not profound compared to SCM and ACM cases.
From this observation,  we argue that the existence of the charges on the particles strengthens the cage effect around certain particle.
Furthermore, this enhanced cage effect causes the crossover behavior in the power law relation.
This crossover behavior observed more profoundly in SCM and ACM can be thought as a distinguishing property of ionic liquids in our model systems.

\begin{figure}[!t]
	\begin{center}
			\includegraphics[width=0.5\textwidth]{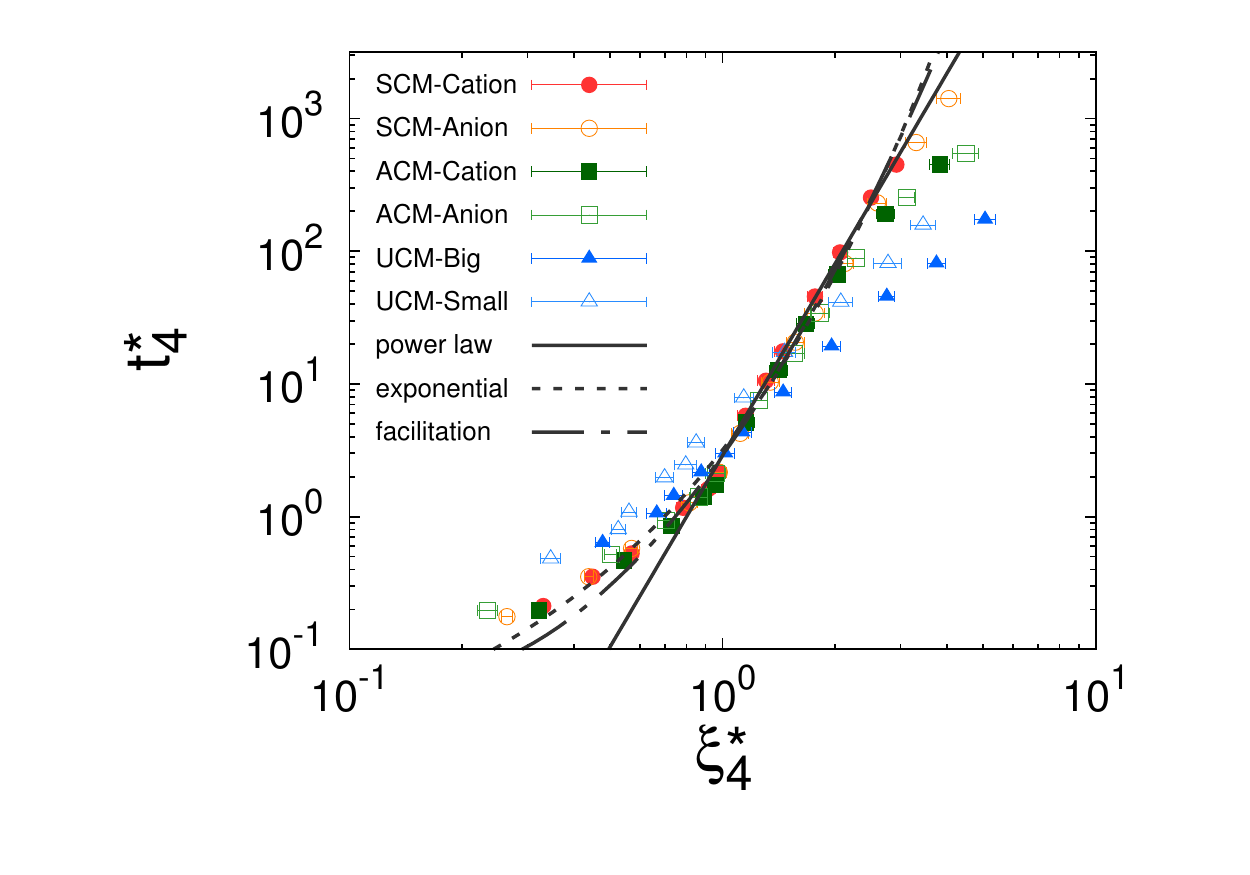}
	\caption{
Comparison of relations between $t_4^*$ and $\xi_4^*$ for SCM, ACM and UCM.
The power law exponent, $\gamma$, of the cation and the anion in SCM and ACM is much larger than the exponent of cation and anion in UCM.
Three different functional form is used to fit the data of the cation in SCM:
(1) $t_4^* \sim {{\xi}_4^*}^{\gamma}$, (2) $t_4^* \sim \text{exp}({\xi_4^*}^z)$, (3) $t_4^* \sim \text{exp}(A(\text{log}({\xi_4^*}/B))^2)$.
      	}
	\label{xi_t4_together}      	
	\end{center}
\end{figure}

When we compare the values of the exponent ${\gamma}$, we find ${\gamma}_{SCM}^{C} \sim 4.8$, ${\gamma}_{ACM}^{C} \sim 4.1$, and ${\gamma}_{UCM}^{C} \sim 2.5$. 
The exponents of the anions are similar to those of the cations.
The exponent value of UCM is similar to the Lennard-Johns mixture studied by La$\check{\text{c}}$evi$\acute{\text{c}}$ et al.\cite{lacevic2003spatially}
It is notable that the lifetime of the dynamic heterogeneity calculated from Ref \citenum{park2015lifetime} has the same exponent, $\zeta_{\text{dh}}$, in all the models in terms of $\tau$, $\tau_{\text{dh}} \sim \tau^{\zeta_{\text{dh}}}$.\cite{park2015lifetime}
Fig.\ref{xi_t4_together} clearly shows that the $t_4^*$ of SCM and ACM increases much faster than the $t_4^*$ of UCM when $\xi_4^*$ is increased.
As we can interpret $\xi_4^*$ as a size of the dynamic cluster,
the same size of the dynamic cluster is preserved longer in time for charged systems.
This is because the local structure of charged system is highly ordered compared to the structure of uncharged system.
Meanwhile, the difference between two charged system, SCM and ACM, is not profound.
In spite of the different charge distributions of the cation,
the crossover behavior is very similar for two systems and
the exponents of the power law between the correlation length and the characteristic time scale of the dynamic heterogeneity have similar values.

Finally, we demonstrate the three fitting schemes to find the relation between $t_4^*$ and $\xi_4^*$.
These fitting schemes have different theoretical bases.
First, mode-coupling theory predicts there is a power law relation between two quantities as we already observed,
\cite{biroli2004diverging,biroli2006inhomogeneous,szamel2008divergent}
$t_4^* \sim {{\xi}_4^*}^{\gamma}$.
Second, the Random-First-Order Transition (RFOT) theory suggests the exponential relation,\cite{kirkpatrick1989scaling,lubchenko2007theory}
$t_4^* \sim \text{exp}({\xi_4^*}^z)$.
Lastly, the view of the facilitation picture suggests the following relation,\cite{keys2011excitations}
$t_4^* \sim \text{exp}(A(\text{log}({\xi_4^*}/B))^2)$.
Fig.\ref{xi_t4_together} shows the functions fitted on the data of the cation of SCM using different schemes.
The lower fitting range is set to be $0.6$ for all functional forms.
It seems that the power law relation is most appropriate to describe the data at long length scale.
The other two functions show similar exponential behavior.
At short length scales, it seems that the exponential function well matches to the data.
However, the data at short length scales are governed by a different physical environment and
it may be a coincidence that the functions agree with the data.
Note that different behavior of short length regime is due to the cage effect of ionic liquid model.
It is notable that Flenner et al. reported that there is a universal behavior of supercooled liquids which is an exponential relation between $t_4^*$ and $\xi_4^*$.\cite{flenner2014universal}

\section{Conclusions}
In the previous works, heterogeneous dynamics has been found in RTILs using the theoretical schemes which are applied to supercooled liquids.
In this study, we find the evidences of heterogeneous dynamics in the simple ionic liquids models.
As the temperature decreases, displacement distribution of the cation and the anion is getting broaden as shown in Fig.\ref{dis_tscan_scm} and Fig.\ref{dis_tscan_scm_ani}.
It seems that the broadening of the distribution reaches its maximum when the time is comparable to the relaxation time of each model system.
Furthermore, the decoupling of the mean exchange time and the mean persistence time is observed.
At low temperatures, the mean persistence time is growing much faster than the mean exchange time as expected from the previous studies.\cite{jung2005dynamical,jeong2010fragility}
From the result, we can infer that the defined excitation events are correlated each other at sufficiently low temperatures,
which is caused by the correlated local motion of the particles.

We adopt the four-point correlation function analysis to study how the local densities are correlated in our ionic liquids model systems.
To quantify the heterogeneous dynamics, the dynamic susceptibility is calculated based on two different time-correlation functions which are $Q(t)/N$ and $F_s(k,t)$.
The time dependence and the temperature dependence of calculated dynamic susceptibility, $\chi_4(t)$ and $\tilde{\chi}_4(t)$, are investigated.
Our results illustrate that the dynamic heterogeneity found in RTILs is transient in time analogous to the situation in the supercooled liquids.

We also successfully extract the dynamic correlation length, $\xi_4^*$, by fitting the dynamic structure factor, $S_4(t)$, into the Ornstein-Zernike equation.
Calculated quantities such as the characteristic time scale of the dynamic heterogeneity, $t_4^*$, the maximum value of the dynamic susceptibility, $\chi_4^*$, and the dynamic correlation length, $\xi_4^*$, are connected via the power law relations at low temperatures.
Interestingly, the crossover behavior around $t_4^* \sim 1$ and $\xi_4^* \sim 1$ is prominent in the charged model, SCM and ACM.
We count this phenomenon on the enhanced cage effect due to the existence of charge.
The crossover behavior and the peak of the dynamic susceptibility in the short time region $t \sim \tau_{\beta}$ are not noticeable in the UCM and in the previous studies on the glassy systems.
Note that, in the Ref \citenum{park2015lifetime}, 
the power law relations between the lifetime of the dynamic heterogeneity, $\tau_{\text{dh}}$, and $\tau$ was investigated.
The power law exponents, $\zeta_{\text{dh}}$, were found to be the same in all the models studied in this work.
In this study, on the other hand, the power law exponents related with the length scale of the dynamic heterogeneity, $\theta$ and $\gamma$, show different values depending on the models.

As we vary the charge distributions on the cation particle, the effect of different charge distributions on the glassy dynamics is observed.
When the result of UCM is compared with the charged model systems, all three models show the heterogeneous dynamics at sufficiently low temperatures.
Even if the particles do not have the charge, the mixture of different shape of particle shows the glassy behavior like other supercooled liquids models.
The existence of charge on the particles mainly affects two aspects of the system.
First, the onset temperature of heterogeneous dynamics increases.
Because the structures of the charged systems are more stable than the structure of UCM at the same temperature,
the dynamics is much slower and the temperature which shows heterogeneous dynamics is much higher.
Second, the cage effect is enhanced.
From the simulation results of dynamic susceptibility and the power law analysis,
we confirm that the alternating local structure of the cations and the anions results strong cage effect.
As a result, we find crossover behavior for the power law relation between the time scale and the length scale of the dynamic heterogeneity.
Furthermore, the comparison between SCM and ACM reveals that the asymmetric charge distribution makes the system more fragile.
However, two models basically show similar behaviors except the onset temperature of the dynamic heterogeneity.

The simple models we used are designed to reveal heterogeneous dynamics which can be observed in RTILs.
Because of their simplicity, a clear comparison of different models is possible and relatively long and large simulation is available compared to the all-atom models of RTILs.
While the models can provide the insight on the role of the charge distributions on the cations,
they are highly coarse-grained models so that the effect of the molecular details are ignored.
In order to extend the models to more realistic systems, detailed molecular structure can be considered.
In the future study,
less coarse-grained model such as 4-atom cation model\cite{jeong2010fragility}
may be used for investigation of the dynamic heterogeneity in RTILs.

\bibliography{main_v2}

\end{document}